\documentclass[a4paper,11pt]{article}
\pdfoutput=1 %

\usepackage{jheppub} 
\usepackage[T1]{fontenc} 
\usepackage[normalem]{ulem}

\usepackage{amsmath,amssymb,graphicx,multirow,xspace,slashed}

\usepackage{xcolor}

\definecolor{darkgreen}{rgb}{0,0.5,0}
\definecolor{goodyellow}{rgb}{0.9,0.7,0}

\def\dd{\text{d}}

\newcommand{\muc}{MuC}

\title{\boldmath Searching for Leptoquarks at Future Muon Colliders }

\author[a]{Pouya Asadi,}
\author[b,c]{Rodolfo Capdevilla,}
\author[d]{Cari Cesarotti,}
\author[d]{Samuel Homiller}

\affiliation[a]{Center for Theoretical Physics, Massachusetts Institute of Technology, Cambridge, MA 02139, USA}
\affiliation[b]{Department of Physics, University of Toronto, Canada}
\affiliation[c]{Perimeter Institute for Theoretical Physics, Waterloo, Ontario, Canada}
\affiliation[d]{Department of Physics, Harvard University, Cambridge, MA 02138, USA}

\emailAdd{pasadi@mit.edu}
\emailAdd{rcapdevilla@perimeterinstitute.ca}
\emailAdd{ccesarotti@g.harvard.edu}
\emailAdd{shomiller@g.harvard.edu}

\preprint{MIT--CTP 5296}

\abstract{
A high energy muon collider can provide new and complementary discovery potential to the LHC or future hadron colliders. 
Leptoquarks are a motivated class of exotic new physics models, with distinct production channels at hadron and lepton machines. 
We study a vector leptoquark model at a muon collider with $\sqrt{s} = 3, 14$ TeV within a set of both UV and phenomenologically motivated flavor scenarios.
We compute which production mechanism has the greatest reach for various values of the leptoquark mass and the coupling between leptoquark and Standard Model fermions. 
We find that we can probe leptoquark masses up to an order of magnitude beyond $\sqrt{s}$ with perturbative couplings. 
Additionally, we can also probe regions of parameter space unavailable to flavor experiments. 
In particular, all of the parameter space of interest to explain recent low-energy anomalies in $B$ meson decays would be covered even by a $\sqrt{s} = 3\,\textrm{TeV}$ collider.
}

\begin{document} 
\maketitle
\flushbottom

\section{Introduction}
\label{sec:intro}

Detection of new physics beyond the Standard Model (SM) has been elusive so far.
While the Large Hadron Collider (LHC) still has the potential to both discover new physics and improve our understanding of the Standard Model, it is becoming increasingly clear that a future high energy collider will be necessary to search for solutions to the shortcomings of the SM. 
Given the abundance of motivated, exotic scenarios for physics beyond the Standard Model (BSM), it is imperative to understand the reach and complementarity of different future collider options for different BSM possibilities. 

A multi-TeV muon collider (\muc) is an especially exciting possibility to succeed the LHC~\cite{Delahaye:2019omf,Shiltsev:2019rfl,Shiltsev:2017tjx,Budker:1969cd,Ankenbrandt:1999cta,Palmer:2014nza,Wang:2015yyh,Alexahin:2018svu,Boscolo:2018ytm,Neuffer:2018yof}.
Such a machine would have several intrinsic benefits beyond that of high-precision lepton colliders and high-energy hadron colliders.
As muons are $\sim\,200$ times heavier than electrons, they produce less synchrotron radiation and can thus be readily accelerated to high center of mass (COM) energies.
At the same time, they preserve many of the advantages of lepton machines: they are not composite objects like the proton, and can thus utilize the full COM energy in their collisions.
Furthermore, as the colliding particles are not strongly interacting, searches with strongly-interacting final states generally have much smaller backgrounds than corresponding channels at hadron colliders. The production of electroweak particles (either in SM or any potentially new ones) therefore constitutes a much larger fraction of the events than at comparable $pp$ colliders. 
Finally, a \muc~ is uniquely capable of testing BSM scenarios that couple to second generation leptons directly, allowing searches for hitherto unexplored flavorful BSM theories whose signals may be suppressed in $e^+e^-$ or hadron collisions.
In this work, we will demonstrate how all of these advantages are brought to bear in searching for particles that couple leptons and quarks at a vertex.

There are several inherent difficulties in working with muons, largely stemming from challenges in their production and their instability. 
It has long been understood that muons produced from meson decays (a design that in principle allows for high luminosity beams) occupy a large volume of phase space and must be cooled in order to create a beam as necessary for high energy collisions~\cite{Parkhomchuk:1983ua,Neuffer:1983jr,Neuffer:1986dg,Kaplan:2014xda,Adey:2015iha}.
Nevertheless, the past few years have seen considerable progress towards realizing high intensity, low emittance muon beams~\cite{Bogomilov:2019kfj,Bonesini:2019dyo,Long:2020wfp,Young:2020wcv}, including an alternative design that circumvents this problem altogether~\cite{Antonelli:2019uoe,Antonelli:2015nla,Boscolo:2018tlu}.
In particular, these studies have led to realistic targets of the integrated luminosity scaling of a high energy muon collider (assuming a 5 year run)~\cite{Delahaye:2019omf}:
\begin{equation}
    L \simeq 10\,\textrm{ab}^{-1} \times \big( \sqrt{s} / 10\,\textrm{TeV} \big)^2.
    \label{eq:lumi}
\end{equation}
We take this scaling as a benchmark to investigate the reach of a $3\,\textrm{TeV}$ ($1\,\textrm{ab}^{-1}$) and a $14\,\textrm{TeV}$ ($20\,\textrm{ab}^{-1}$) machine.

The short lifetime of the muon leads to another experimental challenge: the muon decay products will inevitably interact with the surrounding machinery, giving rise to a significant ``beam-induced background'' (BIB)~\cite{Cummings:2011zz,Kahn:2011zz,Mokhov:2011zzd,Mokhov:2014hza,DiBenedetto:2018cpy,Bartosik:2019dzq,Bartosik:2020xwr,Lucchesi:2020dku}. This effect must be mitigated to study the high energy collisions of interest.
The background can be partially removed by introducing ``shielding nozzles'' in the detector, at the cost of losing detector coverage up to a given opening angle.
In the absence of a realistic detector design optimized for different energies, we will assume that the detector has coverage only up to $|\eta| \leq 2.5$ for all visible particles.\footnote{This estimate on the $\eta$ range is reasonable, as studies indicate that the resolution of soft, forward tracks degrades substantially in large $\eta$ \cite{Bartosik:2020xwr}} Furthermore, we require them to have $p_T \geq 25$~GeV, and assume that the effects of the BIB can otherwise be ignored. 
This assumption is somewhat conservative, as the BIB becomes less problematic at higher energies, and with improvements in timing.

In light of these experimental advances, there has been a proliferation of interest from theoretical physics in muon colliders~\cite{Ali:2021xlw}. 
These studies include both detailed studies of the PDFs that are relevant in muon collisions~\cite{Han:2020uid, Han:2021kes} and estimates of the rates of SM processes and projections for measuring Higgs couplings~\cite{Chiesa:2020awd, Costantini:2020stv, Han:2020pif}.
There have also been numerous studies of the reach for high energy muon colliders to probe various BSM possibilities, including 
new scalars~\cite{Eichten:2013ckl, Chakrabarty:2014pja, Buttazzo:2018qqp, Bandyopadhyay:2020otm, Han:2021udl, Liu:2021jyc}, 
minimal dark matter~\cite{Han:2020uak, Capdevilla:2021fmj, Bottaro:2021srh}, 
explanations for discrepancies in the muon $(g-2)_{\mu}$~\cite{Capdevilla:2020qel, Buttazzo:2020eyl, Capdevilla:2021rwo, Chen:2021rnl, Yin:2020afe} 
or the neutral current $B$-meson anomalies~\cite{Huang:2021nkl, Huang:2021biu}, 
and for new physics encoded in higher-dimensional operators~\cite{DiLuzio:2018jwd,Buttazzo:2020uzc}. 
In this work, we further these theoretical works by studying leptoquarks (LQs), a class of exotic BSM particles with flavorful couplings to SM fermions and unique signals at colliders.

While there are no mediators in the SM that can couple a quark and a lepton at a vertex, there is no reason to believe this is true in the UV theory that completes the SM. 
Such mediators are collectively called LQs.
These particles can emerge from various UV completions of the SM,
and are especially ubiquitous in models with unified gauge interactions \cite{Pati:1973uk, Pati:1974yy, Georgi:1974sy}. 
Certain representations of leptoquarks under the SM gauge group are impossible to obtain in some string theory constructions, so any signal of such a LQ would immediately rule out a number of models~\cite{Ibanez:2012zz}. 
Aside from any UV motivations for their existence, LQs near the weak scale have been well studied in the literature~\cite{Bandyopadhyay:2016oif, Dey:2017ede, Bandyopadhyay:2018syt, Roy:2018nwc, Mandal:2018czf, Chandak:2019iwj, Padhan:2019dcp, Bhaskar:2020kdr, Buonocore:2020erb, Greljo:2020tgv, Bandyopadhyay:2020jez, Crivellin:2021egp}, 
and have been frequently appealed to in order to explain various low energy anomalies e.g. in $B$ meson decays~\cite{Barbieri:2011ci, Barbieri:2015yvd, Fajfer:2015ycq, Bauer:2015knc, Aydemir:2018cbb, Biswas:2018snp, Biswas:2018iak, Aydemir:2019ynb, Hou:2019wiu, Bandyopadhyay:2020klr, Haisch:2020xjd, Cornella:2021sby}.
LQs at the TeV scale manifest in a variety of scenarios.
In supersymmetric theories with $R$-parity violation (RPV), the squarks acquire couplings to leptons and become scalar LQs~\cite{Dawson:1985vr}.
LQs can also emerge near the confinement scale in strongly coupled theories~\cite{Schrempp:1984nj}.
Grand Unified Theories generically have LQs in small representations of the gauge group, and if they belong to a representation that doesn't participate in the GUT symmetry breaking at high scales, their mass may be close to the weak scale~\cite{Frampton:1991ay}.

The unique couplings of LQs to SM fermions leads to a very particular phenomenology at experiments, worthy of in-depth study at both high energy colliders and in low energy observables. There are already various precision low energy experiments probing the signatures of LQs, see ref.~\cite{Dorsner:2016wpm} for a review of various LQ contributions to low energy observables. Such observables always have the highest sensitivity to interactions involving the first generation leptons. Dedicated searches for these LQs are carried out at LHC too \cite{Aaboud:2016qeg, Sirunyan:2017yrk, Sirunyan:2018jdk, Sirunyan:2018ryt, Sirunyan:2018vhk, Sirunyan:2018btu, Sirunyan:2018xtm, Aaboud:2019jcc, Aaboud:2019bye, Aad:2020sgw, Aad:2020iuy, Aad:2020jmj, Sirunyan:2020zbk, Aad:2021rrh, Aad:2021jmg}. In the high energy hadron colliders the production of these particles are mostly dominated by the color production. A thorough study of LQ signals at the LHC is carried out in refs.~\cite{Diaz:2017lit, Bansal:2018eha, Angelescu:2018tyl, Schmaltz:2018nls, Bansal:2018nwp}. 

In contrast to the color production at hadron colliders, the production mechanisms for LQs at a high energy \muc~are sensitive to the electroweak and LQ couplings.
This will allow us to directly probe the couplings of the LQs to muons at an unprecedented precision. 
In this work, as a proof of principle, we will study the reach of a \muc~in the parameter space of the $U_1$ LQ. 
The phenomenology of LQs with different spins or representations of the SM gauge group is expected to be quite similar.

The rest of this paper is structured as follows: in Sec.~\ref{sec:model}, we review the details of the model and discuss its flavor texture. In Sec.~\ref{sec:production} we discuss each of the LQ production modes in turn, and detail our analysis strategy. We collect the constraints from all the production modes in Sec.~\ref{sec:comparison}, and compare to complementary constraints from flavor observables and other future colliders. We conclude in Sec.~\ref{sec:conclusion}. App.~\ref{app:stats} includes further details on our likelihood analysis, while App.~\ref{app:kappa} provides further details on the effect of modified LQ couplings to SM gauge bosons on the reach of a \muc{} in their parameter space.

\section{Model}
\label{sec:model}

The goal of this work is to understand the phenomenolgy of LQs at a high energy muon collider.
An LQ refers to any particle that couples leptons and quarks at a vertex, and they can therefore have a variety of properties: they can be scalars or vectors, and can have several possible representations under the electroweak gauge group~\cite{Buchmuller:1986zs}. 
As mentioned in the introduction, LQs arise in a variety of BSM scenarios, including GUTs, supersymmetric theories, and composite models.
All of these scenarios carry additional complications, and generally lead to other new states and signatures near the LQ mass.
However, the signatures of the LQs themselves in all of these scenarios are closely related. 
We will therefore adopt a simple, phenomenological parameterization of LQ interactions to understand the relevant features at a muon collider.
In this work, we thus focus on one well-motivated LQ, referred to as $U_1$ in the notation of ref.~\cite{Dorsner:2016wpm}.
With small modifications, our results can easily be adapted to other LQs at a high energy \muc.\footnote{See ref.~\cite{Huang:2021biu} for a recent study of the bounds from a subset of the signals that we study on the scalar LQ $S_1$.}

The $U_1$ is a vector LQ in the $(\mathbf{3}, \mathbf{1}, 2/3)$ representation of the SM gauge group. 
It was the first LQ to appear in the literature, as it emerges in the spectrum of Pati-Salam grand-unified theories (GUTs)~\cite{Pati:1973uk, Pati:1974yy}.
It is a genuine LQ --- it has no di-quark interactions --- and therefore does not give rise to proton decay. 
Simplified models with a $U_1$ LQ have been studied as a particularly intriguing solution to the recent $B$-meson anomalies referenced in the introduction.
We will discuss this connection more in Sec.~\ref{sec:comparison}.
Depending on the UV-completion, $U_1$ can either be the massive remnant of a spontaneously broken gauge symmetry~\cite{DiLuzio:2017vat, Assad:2017iib, Calibbi:2017qbu, Bordone:2017bld} or a composite of some strongly interacting sector~\cite{Barbieri:2017tuq, Blanke:2018sro}.
In either case, the signatures of the LQ itself can still be understood within the phenomenological framework discussed below, and in that context, our results can be taken as a roadmap for future studies of the different UV scenarios.

Following \cite{Baker:2019sli}, the Lagrangian can be written as follows: 
\begin{align}\label{eq:u1_lag}
\mathcal{L}_{U_1} & =  
-\frac{1}{2} U_{1\,\mu\nu}^{\dagger} U_{1}^{\mu\nu} 
+ m_{\textrm{LQ}}^2 U_{1\,\mu}^{\dagger} U_1^{\mu}
-i g_s (1 - \kappa_U) U_{1\,\mu}^{\dagger} T^a U_{1\, \nu} G^{a\, \mu\nu} \nonumber \\ 
&
- i g_Y \frac{2}{3}(1 - \tilde{\kappa}_U) U_{1\,\mu}^{\dagger} U_{1\, \nu} B^{\mu\nu} 
+ \frac{g_U}{\sqrt{2}} \bigg[ 
U_1^{\mu}\big( 
\beta^{ij}_L \bar{Q}_L^i \gamma_{\mu} L_L^j 
+ \beta^{ij}_R \bar{d}_R^i \gamma_{\mu} e_R^j 
\big) + \textrm{h.c.}
\bigg] ,
\end{align}
where $U_{1\,\mu\nu} = D_{\mu}U_{1\,\nu} - D_{\nu}U_{1\,\mu}$ with $D_{\mu}$ the covariant derivative.
Here, $i, j$ are flavor indices of the quarks and leptons respectively, and $T^a$ the $SU(3)$ generators.
In Eq.~\eqref{eq:u1_lag}, we include potential non-minimal interactions with the SM gauge fields, parameterized by $\kappa_U, \tilde{\kappa}_U$. 
In models where $U_1$ is a fundamental gauge boson we have
$\kappa_U = \tilde{\kappa}_U = 0$~\cite{Rizzo:1996ry}.
In strongly interacting theories, however, they can take different values, including the so-called ``minimal coupling scenario''~\cite{CiezaMontalvo:1992bs, Blumlein:1992ej}, between the leptoquarks and gauge fields, where $\kappa_{U} = \tilde{\kappa}_U = 1$.
We have also factored out an overall coupling of the LQs to matter fields, $g_U$, with the relative couplings dictated by flavor spurions $\beta_L^{ij}$, $\beta_R^{ij}$. 
Throughout this work we set $g_U=1$, as it changes only the overall strength of the interactions governed by the spurions $\beta_{L,R}^{ij}$.
We neglect any couplings to hypothetical right-handed neutrinos, which would couple the neutrinos to the LQ and right-handed up-type quark fields.

For definiteness, we will work in the flavor basis in which the fermion doublets are aligned in the down-quark sector, so that
\begin{equation}
Q_L^i = \begin{pmatrix} V_{ji}^* u_L^j \\ d_L^i \end{pmatrix}, 
\qquad
L_L^i = \begin{pmatrix} \nu_L^i \\ e_L^i \end{pmatrix}.
\end{equation}

In principle, $\beta_{L}$ and $\beta_R$ are complex, $3\times 3$ matrices with unspecified coefficients.
The couplings to first generation quarks and leptons are highly constrained by various low-energy experiments (see ref.~\cite{Dorsner:2016wpm} for a review), so we will consider the following Ansatz:
\begin{equation}
\label{eq:betas}
\beta_L^{ij} = \begin{pmatrix}
0 & 0 & 0 \\
0 & \beta_L^{22} & \beta_L^{23} \\
0 & \beta_L^{32} & \beta_L^{33}
\end{pmatrix},
\qquad
\beta_R^{ij} = 0.
\end{equation}
Non-zero right-handed couplings are necessitated by some UV completions~\cite{Pati:1973uk,Pati:1974yy}, but our phenomenological results are insensitive to the chirality of the couplings.
Therefore we keep only the left-handed couplings to quarks and leptons for simplicity, setting all $\beta_R^{ij} = 0$.
Only the left-handed couplings are required to explain the recent $B$-meson anomalies, as we will discuss more in Sec.~\ref{sec:comparison}.
The collider observables we consider are also insensitive to any phases in the $\beta_L^{ij}$, so we will further assume all the components are real, for simplicity.

\begin{table}[]
    \renewcommand{\arraystretch}{1.2}
    \centering
    \begin{tabular}{|c|c|c|c|c|}
    \hline
        Scenarios & 1 &  2 & 3 & 4\\
        \hline
        $\left(\beta_L^{22},~\beta_L^{23},~\beta_L^{33} \right)=$ & $\left(0,~0,~0\right)$ & $\left(\beta_L^{32},~0,~0\right)$ & $\left(0,~0.1,~1\right)$ & $\left(\beta_L^{32},~0.1,~1\right)$\\
        \hline
    \end{tabular}
    \caption{Different flavor structure scenarios we consider in this work. The LQ mass $m_{{\rm LQ}}$ and its Yukawa coupling to $b\bar{\mu}$ ($\beta_L^{32}$) are treated as free parameters in our study. We fix the other couplings to the values indicated in this table to study the effect of the $\beta_L$-matrix flavor structure.}
    \label{tab:flavor_structure}
\end{table}

The $\beta_L^{ij}$ dictate the phenomenology of the LQ at a muon collider via the production, decay, and relevant backgrounds.
The production of LQs at a muon collider depends in particular on the direct couplings to muons --- either $\beta_L^{32}$ or $\beta_L^{22}$.   
We will consider $\beta_L^{32}$ as a free parameter throughout, and fix the other $\beta_L^{ij}$ to different values in several scenarios given in Tab.~\ref{tab:flavor_structure}.
These structures are somewhat motivated by a $U(2)^5$ symmetry breaking pattern, which allows for a potentially large $\beta_L^{33}$ coupling and treats the off-diagonal $\beta_L^{32}$ as a spurion~\cite{Barbieri:2011ci}.
The four scenarios are primarily chosen, though, to give a representative picture of the phenomenology at a muon collider in different flavor structures.

The values of the $\beta_L^{ij}$ also dictate the branching ratios of the $U_1$. Depending on the scenario, either $b\bar{\mu}$ (scenarios 1 and 2) or $b\bar{\tau}$ (scenarios 3 and 4) will be most relevant decay channel of the LQ.
It is worth emphasizing here that since we consider left-handed couplings of the LQ, the LQ coupling to down-type quarks and charged leptons will always come with a corresponding coupling to up-type quarks and neutrinos:
\begin{equation}
\mathcal{L}\supset \frac{g_U}{\sqrt{2}} \beta_L^{32} \big[ \bar{b}_L \gamma_{\alpha} \mu + \big( V^*_{ub} \bar{u}_L + V^*_{cb} \bar{c}_L +V^*_{tb} \bar{t}_L \big) \gamma_{\alpha} \nu_{\mu}
 \big]U_1^{\alpha} .
\label{eq:U1top}
\end{equation}
This means that even in a minimal setup with $\beta_L^{22},\beta_L^{23},\beta_L^{33}=0$, the LQ will have two dominant decay channels: $U_1\to b \bar{\mu}$ and $U_1 \to t \bar{\nu}_\mu$, with roughly equal branching ratios.
Additional channels such as $U_1 \to u \bar{\nu}_\mu$ or $U_1 \to c \bar{\nu}_\mu$ are suppressed by small CKM factors and are subdominant.
In the remainder of this work, we will focus on the final states with a charged lepton and down-type quark as these should be the easiest to identify, but we note that final states with a top and large missing momentum could also be interesting, as top jets will be relatively easy to identify at a muon collider due to the smaller hadronic backgrounds.

Finally, we note that if all the $\beta_L^{ij}$ are small (as is possible in our flavor scenarios 1 and 2 when $\beta_{L}^{ij} \ll 0.1$), the total LQ decay width can be smaller than the QCD scale ($\Lambda_{\textrm{QCD}} \sim 200\,\textrm{MeV}$).
At these timescales, hadronization effects become important and can significantly impact the decay rates and phenomenology, leading distinct event topology such as displaced tracks (see e.g., the discussion in ref.~\cite{Hiller:2018wbv}). A detailed treatment of these effects is beyond the scope of this work. We will make note of the regimes in parameter space where these hadronization effects are relevant and emphasize that our results should be interpreted with caution in those regimes.

\section{Production Modes}
\label{sec:production}

This section is devoted to understanding the three most important production channels of a LQ at a muon collider.
In ref.~\cite{Schmaltz:2018nls}, it was demonstrated that the phenomenology of LQs at hadron colliders can be classified into pair production (PP), single production (SP), and the LQ interference with the SM Drell-Yan process (DY)~\cite{Bansal:2018eha,Raj:2016aky}.
The same classification of LQ production topologies applies to a muon collider as well, though the couplings dictating the production are quite different than at the LHC.
In the following subsections, we review each of these topologies in turn, with an emphasis on the different parts of parameter space they are sensitive to.
We devise a simple set of cuts for each channel and calculate the reach of a \muc{} with different COM energies for probing the parameter space of the $U_1$ LQ described in the previous section.

For both single and pair production, depending on which of the flavor scenarios in Tab.~\ref{tab:flavor_structure} we are studying, the final states most useful in searching for LQs will be different. 
In scenarios 1 and 2, the most identifiable final state will be $U_1 \to b\bar{\mu}$, and we will always focus on this decay channel unless otherwise noted.
In flavor scenarios 3 and 4, however, the large $\beta_L^{33}$ coupling means that the $U_1 \to b\bar{\tau}$ decay mode always has an appreciable branching ratio, while $\textrm{BR}(U_1 \to b\bar{\mu}) \ll 1\%$ already for $\beta_L^{32} < 0.1$. We will therefore consider only $\tau$ final states for scenarios 3 and 4, though a more detailed analysis could in principle include additional information from other final states as well. As shown below, different final states have dramatically different backgrounds. We assume a fully efficient tagging for all different final state particles.  

Furthermore, for the rest of this section, we will assume $\kappa_U,\tilde{\kappa}_U = 0$.
This is the case in UV scenarios where the $U_1$ arises as the gauge boson of some larger symmetry group, and it allows for large, direct couplings of the LQ to gauge fields. These interactions are important for both the SP and PP channels. As mentioned in Sec.~\ref{sec:model}, 
nonzero values of $\kappa_U, \tilde{\kappa}_U$ appear in strongly-interacting UV completions of the $U_1$ LQ. We will briefly revisit these scenarios in App.~\ref{app:kappa}. This mostly influences the PP bounds that has only mild dependence on $\tilde{\kappa}_U$.

We simulate the signal of each channel using \texttt{MadGraph5}~\cite{Alwall:2011uj}, and a modified version of the $U_1$ LQ model file introduced in ref.~\cite{Baker:2019sli}. 
We use the pipeline developed in refs.~\cite{Asadi:2017qon, Perelstein:2009} to scan over the parameter space of the model.
We use standard statistical analyses described in App.~\ref{app:stats} to calculate the potential discovery reach and exclusion bounds on the parameter space of the model.

As the muon collider program is in its infancy, a number of experimental and theoretical studies are required before a reliable estimate of the systematic uncertainties can be included. We therefore neglect all sources of systematic uncertainties in what follows.
In this sense, our results can be interpreted as a ``best-case'' scenario for the reach of a \muc in probing the parameter space of our simplified LQ model.
Once the systematic uncertainties are determined, it will be straightforward to include them in our analysis, see refs.~\cite{Cowan:1998ji,Cowan:2010js} for prescriptions on how to include the systematic uncertainties in a calculation like ours.

\subsection{Leptoquark Pair Production}
\label{subsec:PP}

\begin{figure}
\centering
    \includegraphics[width=0.3\linewidth]{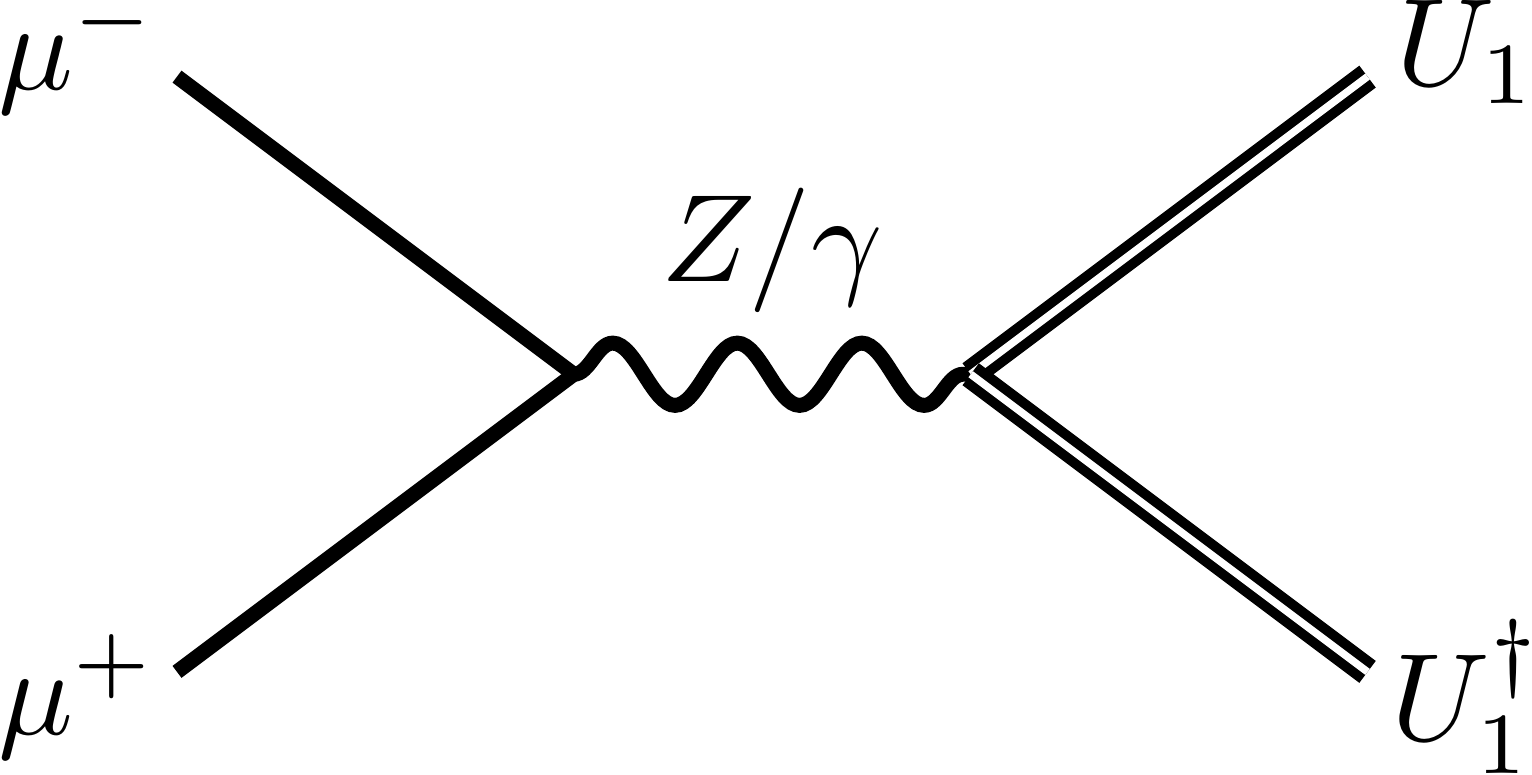}
\hspace{1cm}
\includegraphics[width=0.3\linewidth]{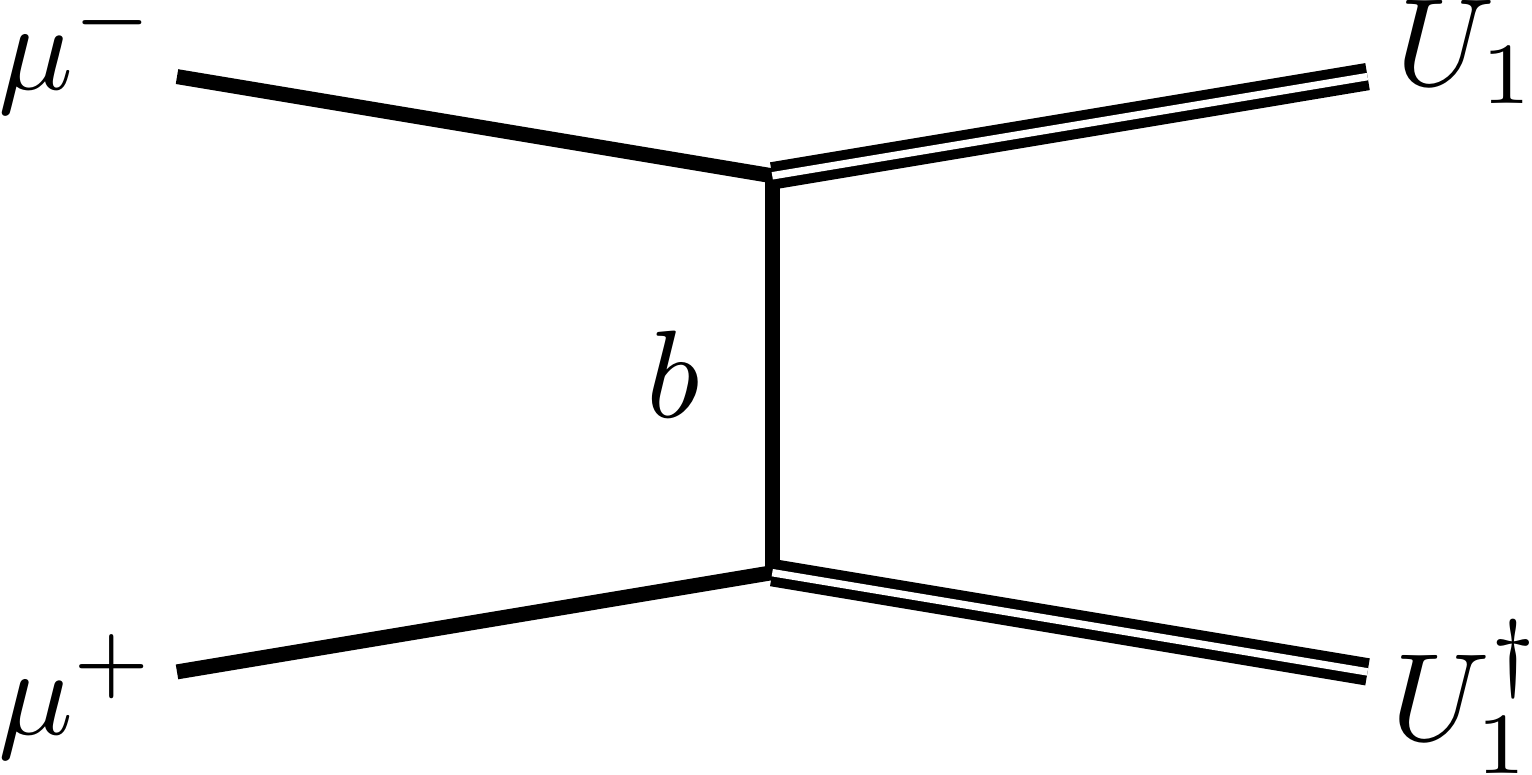} \\
\vskip 0.75cm
\includegraphics[width=0.3\linewidth]{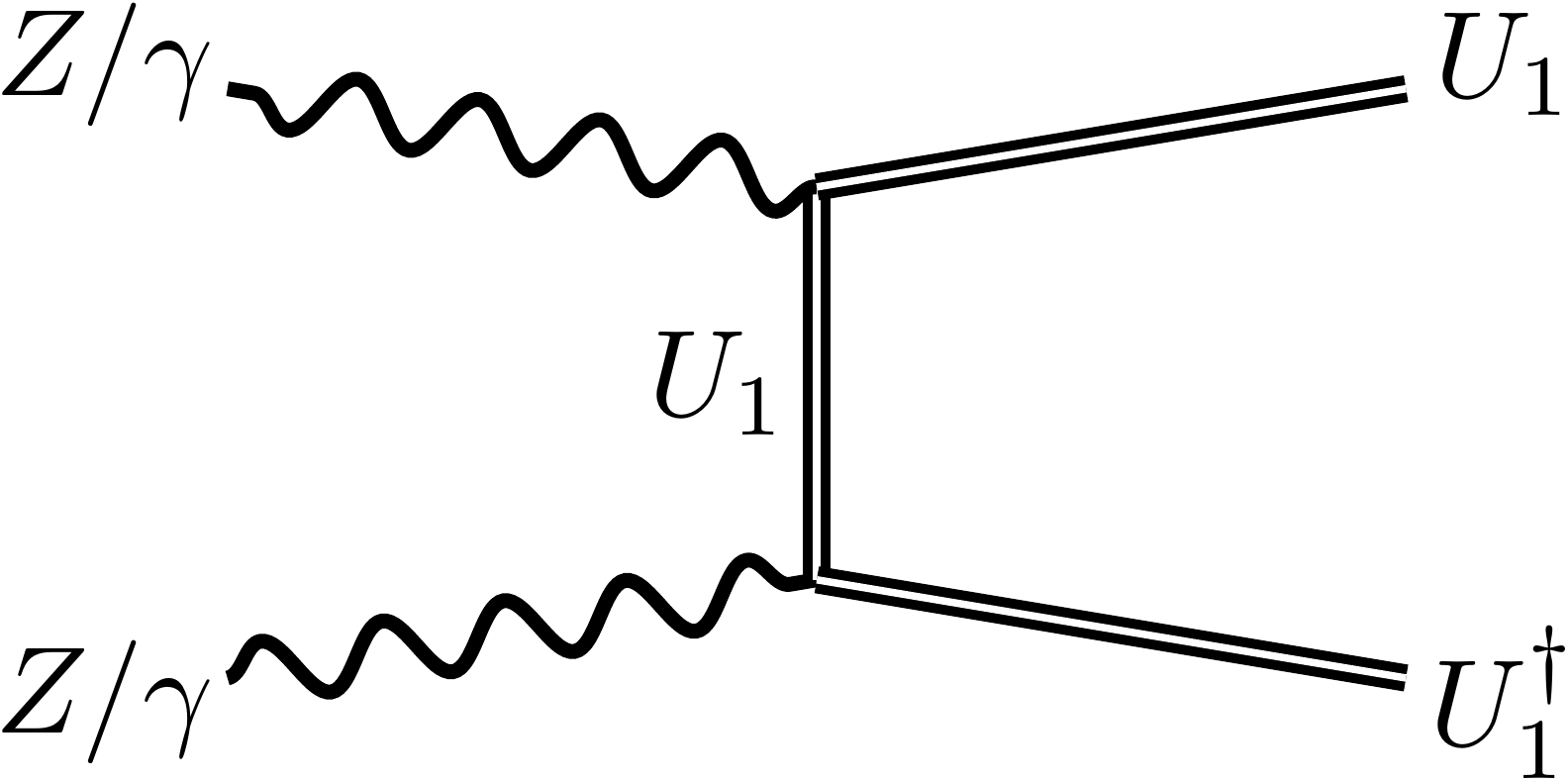}
\hspace{1cm}
\includegraphics[width=0.3\linewidth]{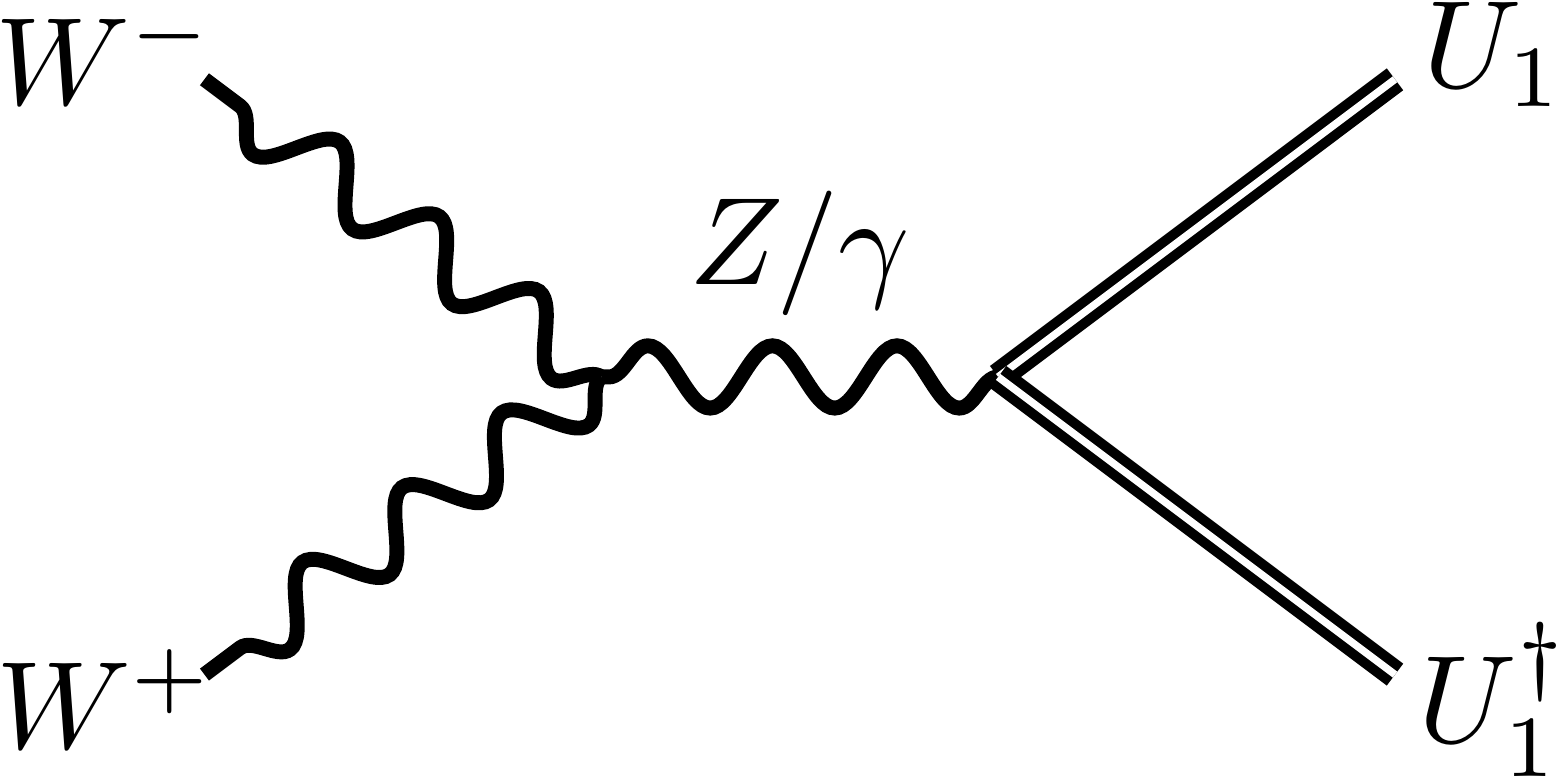}
\caption{
Leading diagrams giving rise to LQ pair production at \muc{}.
The top row shows direct pair production from muon collisions, while the bottom row shows possible contributions from VBF-like diagrams, where the gauge bosons are to be understood as arising collinear radiation from the radiation beam and the remanant particle is unobserved. (See Sec.~\ref{subsec:SP} for more details.) Except for the top-right diagram, all the other ones only depend on the electroweak gauge couplings.
}\label{fig:diag_pairprod}
\end{figure}

\begin{figure}
\centering
\includegraphics[width=0.3\linewidth]{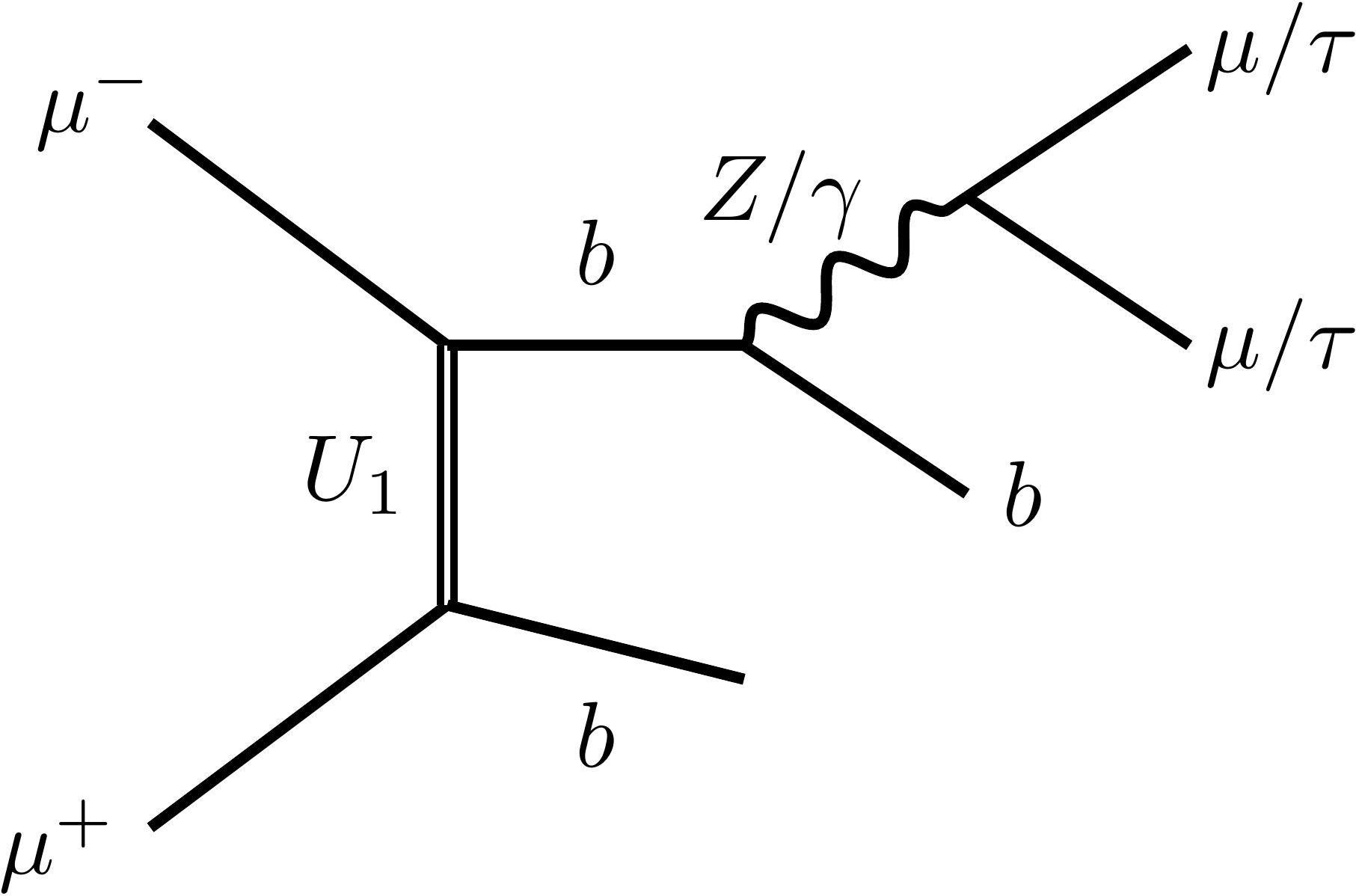}
~
\includegraphics[width=0.3\linewidth]{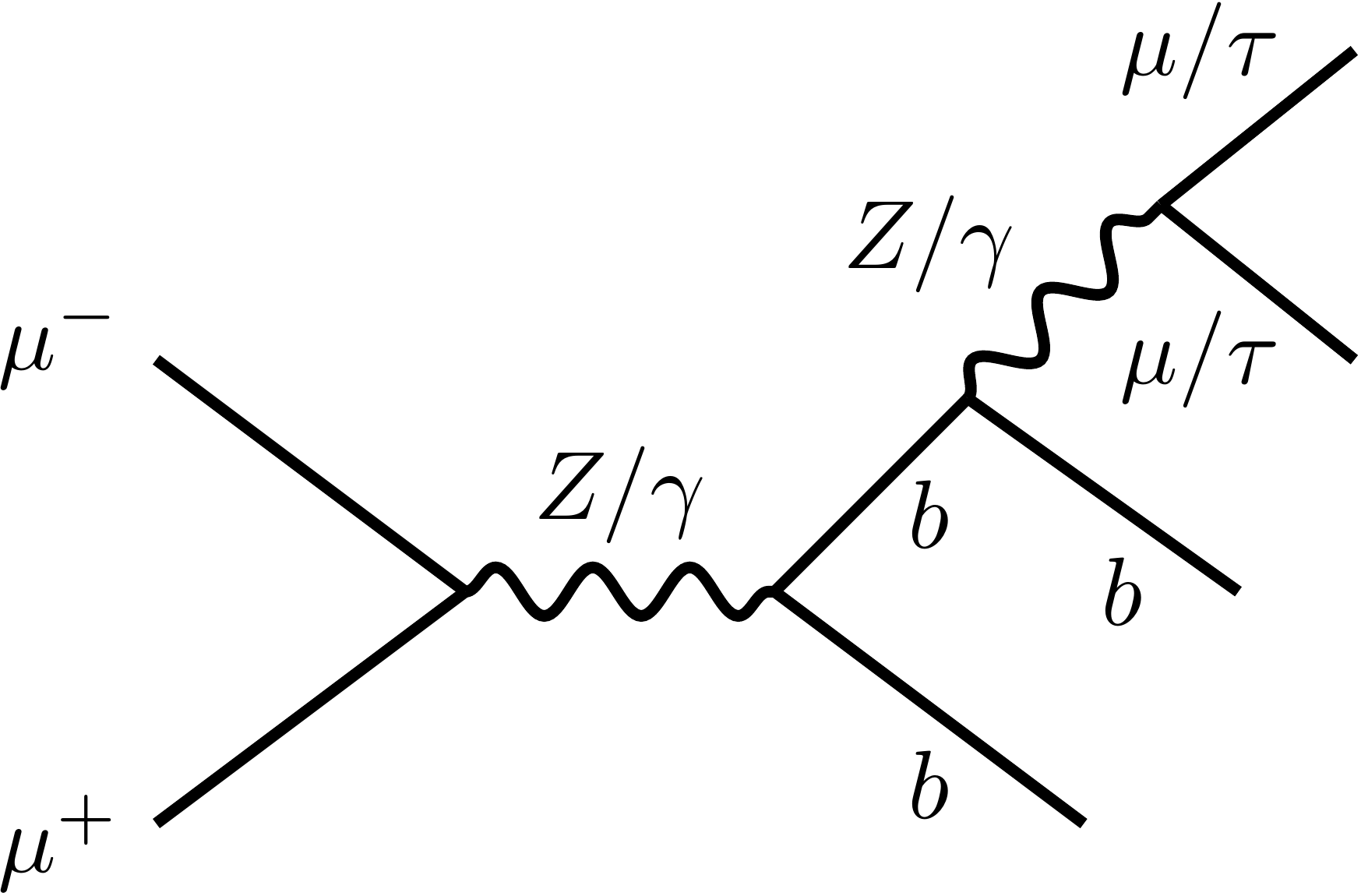}
~
\includegraphics[width=0.3\linewidth]{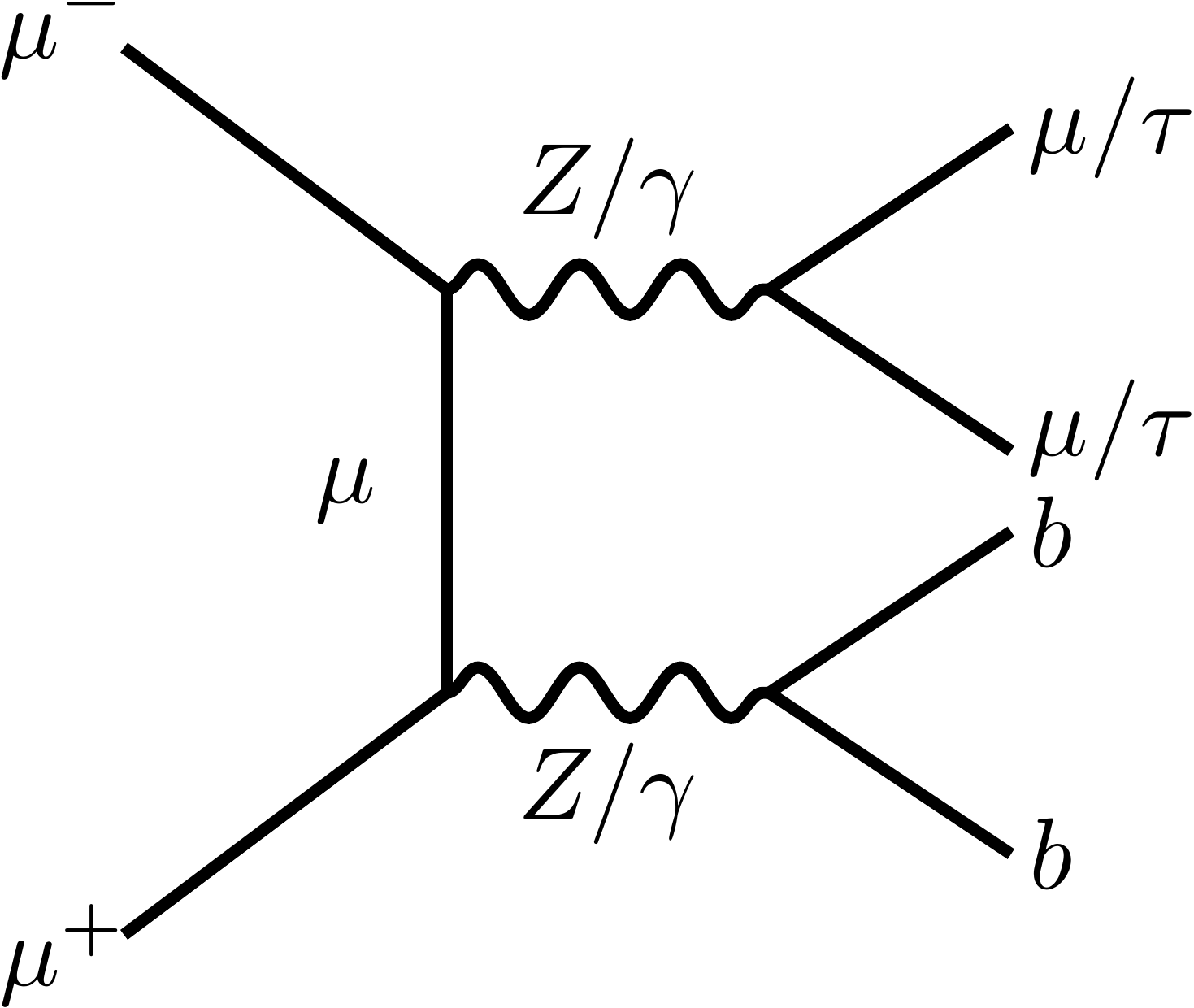}
\caption{
An example ``barking dog'' diagram with one intermediate LQ leading to the same final states as in Fig.~\ref{fig:diag_pairprod} (left), as well as representative diagrams leading to the same final state as the PP signal in the SM (center and right). The distinct topologies of SM and LQ contribution to these final states gives rise to different kinematic observables that we can cut on.
}\label{fig:diag_pairprod_bkg}
\end{figure}

If the LQ is not too heavy, it can be directly pair produced from muon collisions. A priori, PP of a LQ occurs through $s$-channel $\gamma/Z$ exchange, vector boson fusion (VBF) processes, or $t$-channel exchange of a quark (depending on its couplings to muons) as depicted in Fig.~\ref{fig:diag_pairprod}.
However, there are other processes that lead to the same final states and include contributions from LQs, e.g. the ``barking dog'' topology, which only involves one LQ, as in the left panel of Fig.~\ref{fig:diag_pairprod_bkg}. For this reason, in order to be inclusive, here we define ``pair production'' as all contributions involving LQs that lead to a final state with two $b$-jets and two leptons (either muons or taus, depending on the flavor scenario).

Unlike the SP or DY channels discussed below, LQs can be pair produced via their electroweak interactions, even if they lack any direct couplings to muons.
As we will show, for $\beta^{i2} \lesssim 0.2$,
the PP cross section depends only on the LQ mass.
The PP mode is also particularly distinctive at a muon collider, as it leads to two quark-lepton pairs produced back-to-back in the collider for a broad range of LQ masses.

The backgrounds to LQ pair production at a muon collider arise entirely from SM electroweak production of lepton or jet pairs, see Fig.~\ref{fig:diag_pairprod_bkg}. 
The largest SM rates arise when a jet or lepton pair is near the $Z$-pole, but these can be substantially suppressed with a simple cut on the invariant mass. 
For scenarios where the signal requires muons in the final states, the SM $bb\mu\mu$ background also receives significant contributions from topologies where the $b$-pair is produced via fusion of vector bosons radiated off the incoming muons, which continue in the forward direction. 
While the resulting muon pair will be well-separated -- mimicking the signal -- this background is still well-mitigated by requiring the muons to be in the central region ($|\eta_{\ell}| < 2.5$), and by requiring a large invariant mass for the $b$-pair.
Additional backgrounds with missing forward particles can be removed by cutting events where the four visible particles do not have total energy $\sim\sqrt{s}$, so we do not consider them here.

\begin{figure}
\centering
\includegraphics[width=0.48\linewidth]{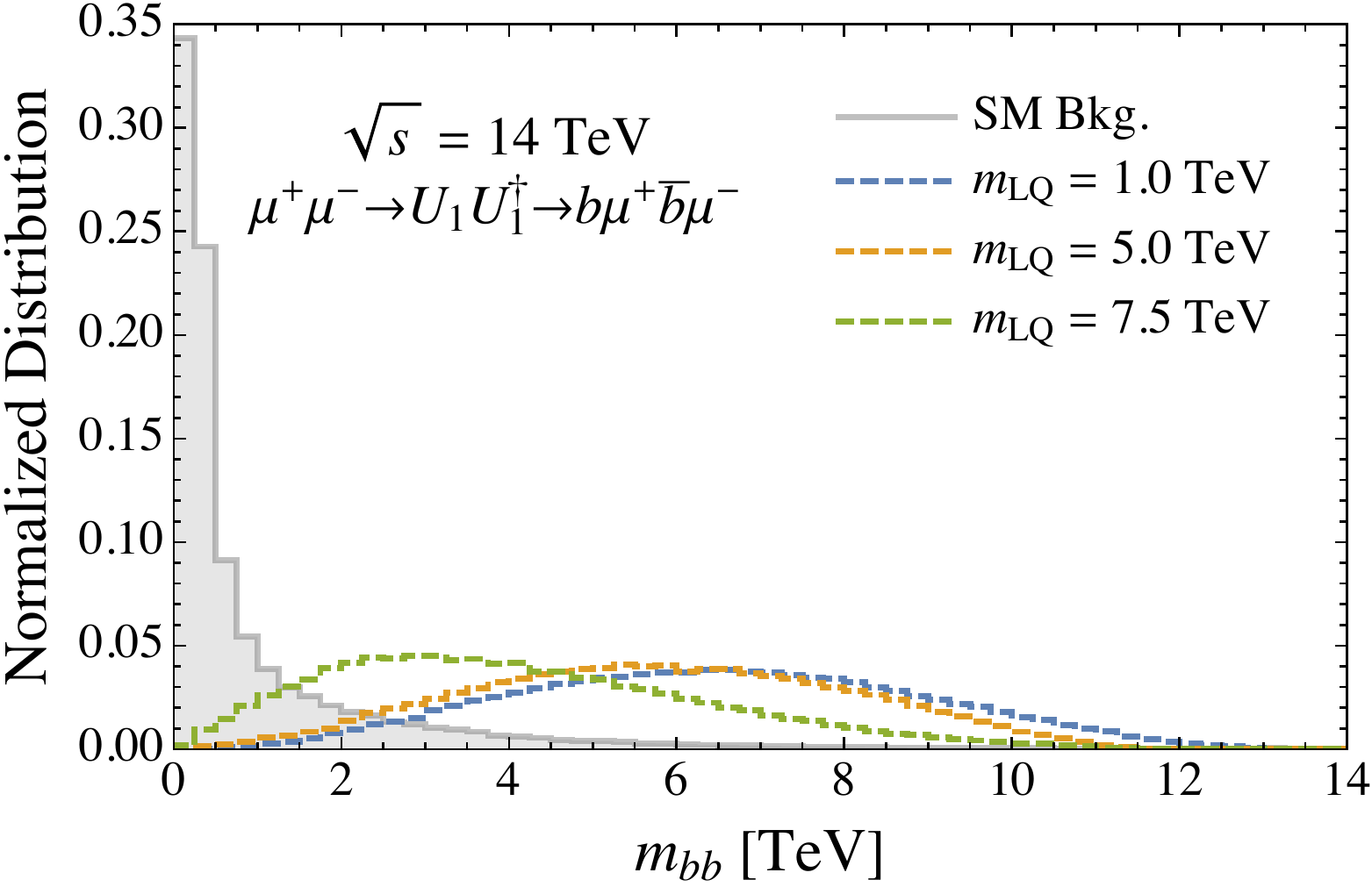}
~
\includegraphics[width=0.48\linewidth]{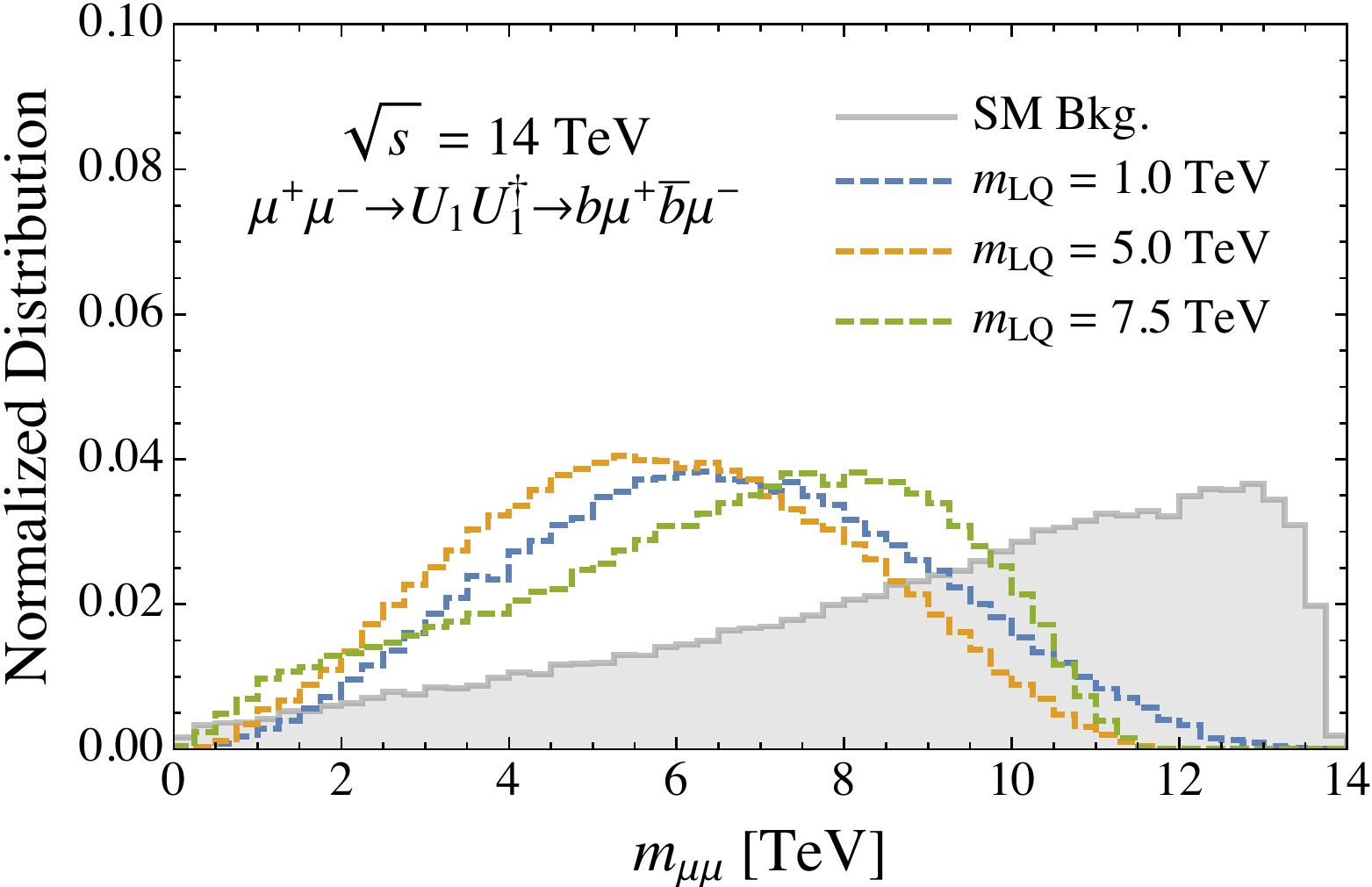}
\vskip 0.2cm
\includegraphics[width=0.48\linewidth]{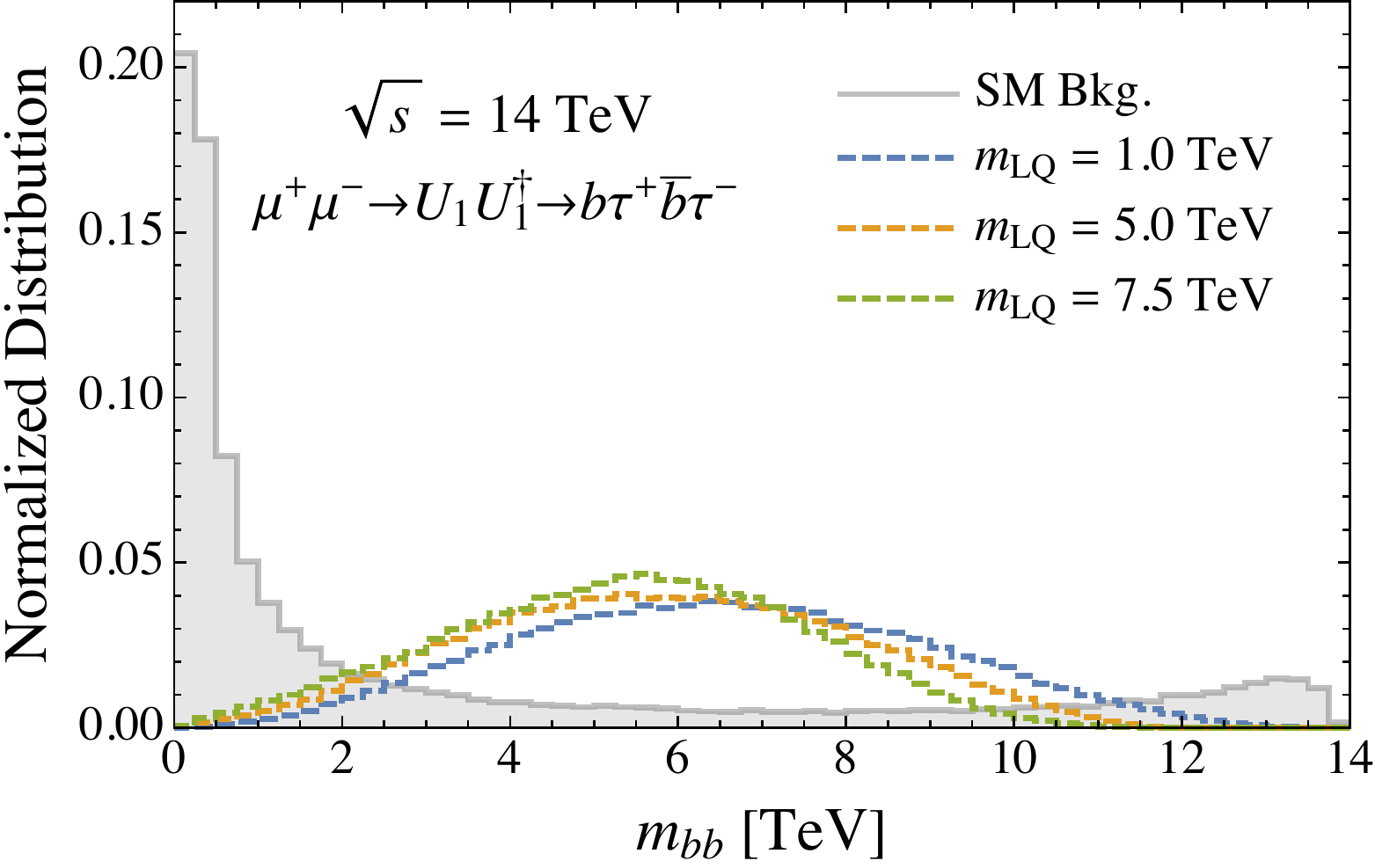}
~
\includegraphics[width=0.48\linewidth]{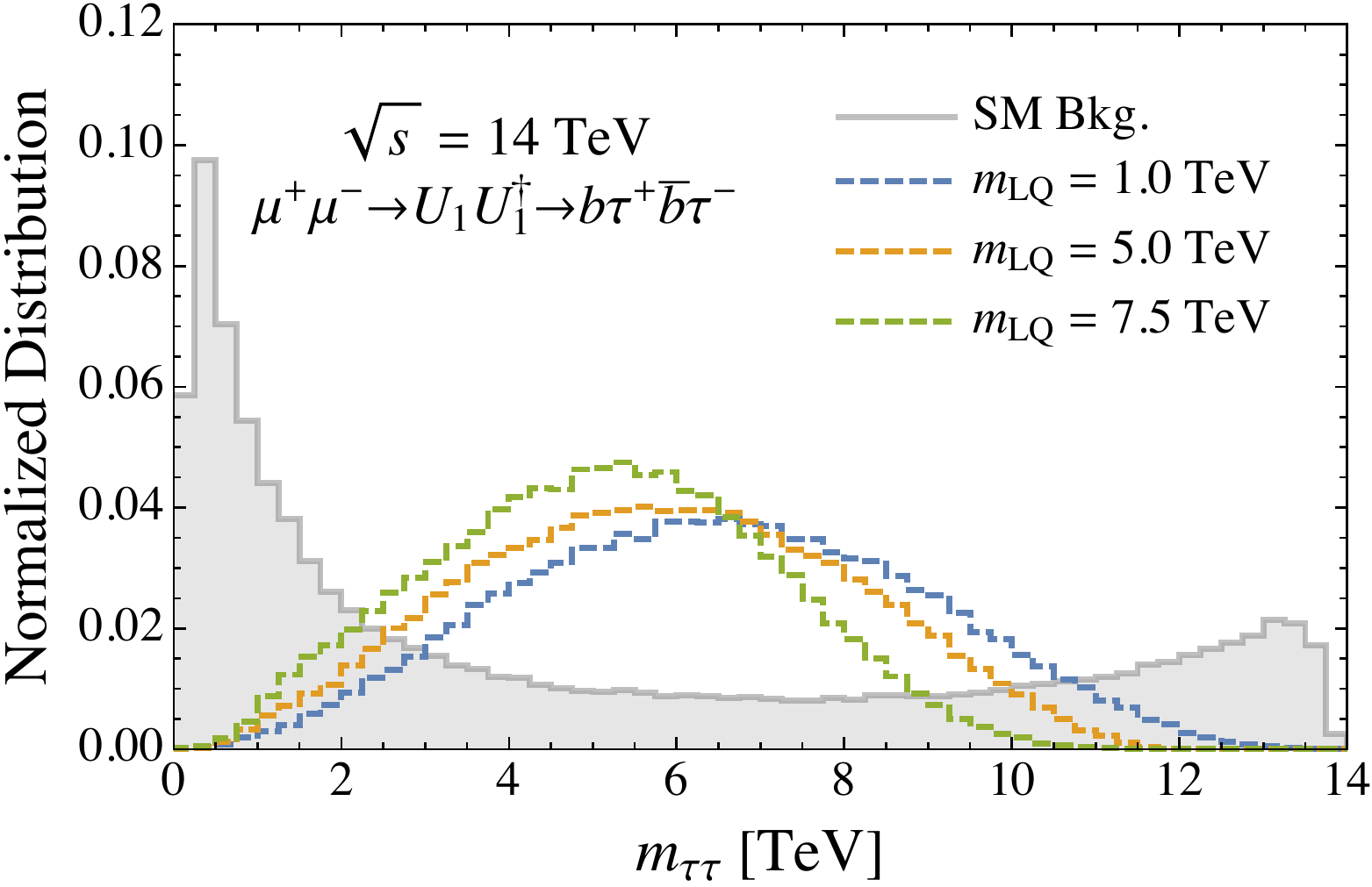}
\caption{
Normalized distributions of the invariant mass for particle-antiparticle pairs in the SM background (gray shaded) and for LQ PP signals. We use $\beta_L^{32}=0.1$ in all the figures. The left two panels show the $m_{bb}$ distribution, while the right two panels show the $m_{\mu\mu}$ (top) and $m_{\tau\tau}$ (bottom) distributions. The upper two panels correspond to flavor scenario 1, with $bb\mu \mu$ final states, while the bottom two panels correspond to flavor scenario 3, with $b b\tau \tau$ final states. The histograms motivate cuts on $m_{bb}$ for improving the signal-to-background ratio; see the text for further details.
}
\label{fig:pairprod_kinematics}
\end{figure}

In Fig.~\ref{fig:pairprod_kinematics} we plot the invariant masses of the various particle-antiparticle pairs for the signal and background in $bb\mu\mu$ and $bb\tau\tau$ final states, both in the SM (gray) as well as for the LQ signal with several different choices of $m_{\mathrm{LQ}}$.
As is clear from the figure, the SM production prefers the jet pair to be highly collimated, and the distribution falls very rapidly as $m_{bb}$ increases. The signal, on the other hand, peaks at roughly $\sqrt{s}/2$, with a slight downward shift for heavier LQs.
The invariant masses of the lepton pair in the signal look quite similar.
While the SM distribution of $m_{\tau\tau}$ in $bb\tau\tau$ production is almost identical to the $m_{bb}$ distribution, the SM $m_{\mu\mu}$ distribution peaks at high invariant masses, due to the VBF-like topology discussed above.

Motivated by this behavior, we impose a cut on the invariant mass of the $b$-quark pair $m_{bb}$ as a function of the collider energy:
\begin{equation}\label{eq:pairprod_cuts}
    m_{bb} > 0.5,~ 2.0~\textrm{TeV} 
    \quad \text{for} \quad 
    \sqrt{s} = 3,~ 14~\textrm{TeV},
\end{equation}
along with a cut on $m_{\mu\mu} > 150\,\textrm{GeV}$ to eliminate the $Z$-pole, this reduces the expected number of background events to $\mathcal{O}(30)$ for each of the integrated luminosity scenarios considered above, while retaining essentially all of the signal. 
In the flavor scenarios 3 and 4, where we utilize $bb\tau\tau$ final states, we impose the same cut in Eq.~\eqref{eq:pairprod_cuts} on $m_{\tau\tau}$, which further reduces the SM $bb\tau\tau$ background to $< 1 $ event in all scenarios.
Note that these estimates make no use of the resonant behavior expected in the $m_{b\mu}$ (or $m_{b\tau}$) distribution, so they are relatively insensitive to energy/momentum resolution of the detector.

\begin{figure}
\centering
\includegraphics[width=0.6\linewidth]{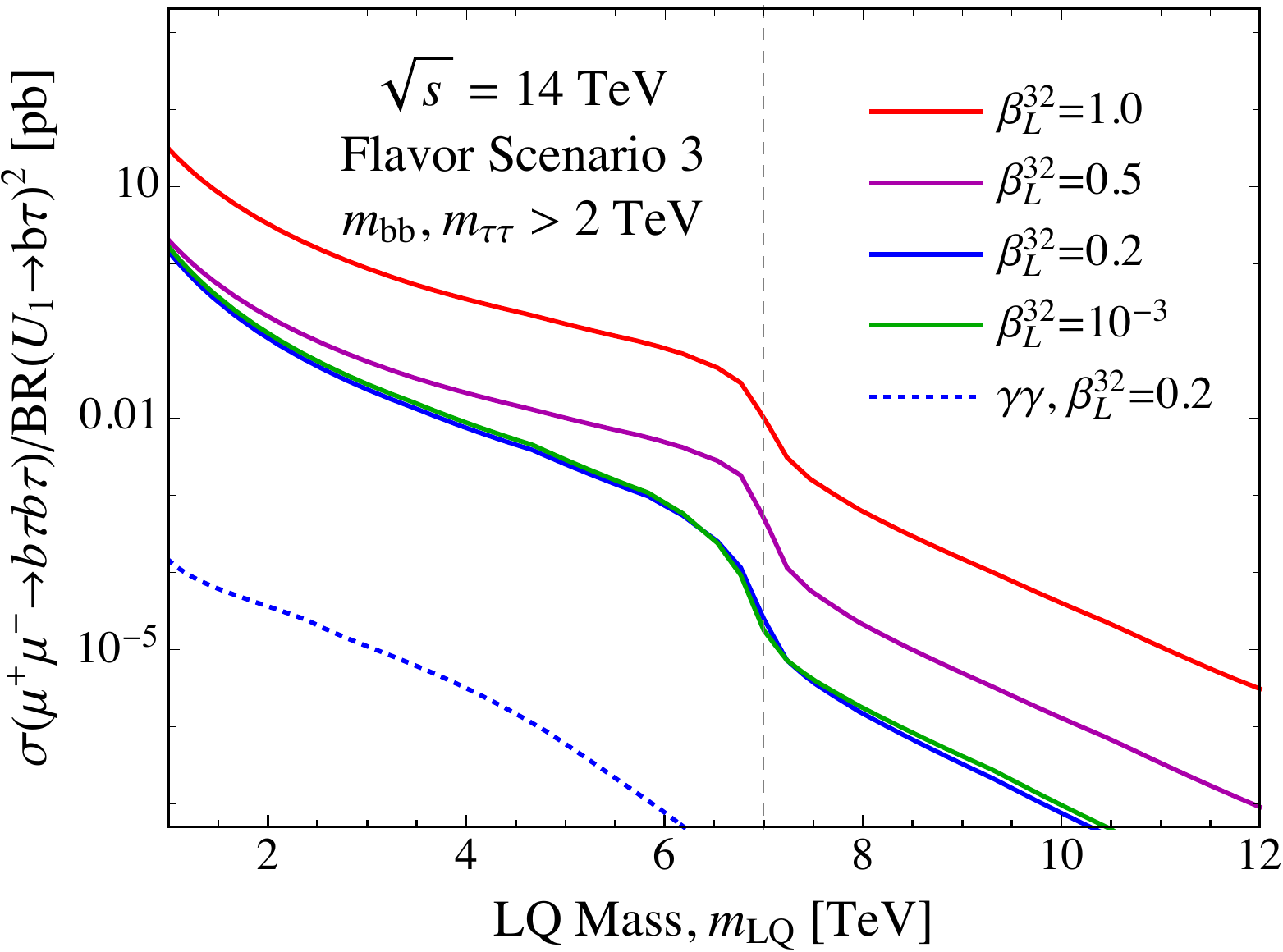}
\caption{
Plot of the PP cross section at $\sqrt{s} = 14\,\textrm{TeV}$ as a function of the LQ mass for several values of $\beta_L^{32}$, normalized by the branching ratio of the LQ. 
The solid curves show the direct $\mu^+\mu^-$ cross section while the dashed curve shows the VBF-induced process computed using the effective photon approximation (see the diagrams on the bottom row of Fig.~\ref{fig:diag_pairprod}). 
The dashed gray line indicates the $\sqrt{s}/2$ threshold. For couplings $\beta_L^{32} \lesssim 0.2$ the production cross section is dominated by the electroweak production for all different masses.
}\label{fig:pairprod_csx}
\end{figure}

In Fig.~\ref{fig:pairprod_csx}, we show the PP cross section after the cuts described above for $\sqrt{s} = 14\,\textrm{TeV}$ as a function of the LQ mass, factoring out the branching ratios of the LQs.
The solid curves show direct PP cross section for several values of $\beta_L^{32}$, assuming the other $\beta_L^{ij}$ vanish.
The dashed blue curve shows the VBF-induced mode, from the bottom left diagram in Fig.~\ref{fig:diag_pairprod}.\footnote{For details on how the photon initial-states are treated, see Sec.~\ref{subsec:SP}.} 
The $\gamma\gamma \to U_1 U_1^{\dagger}$ production is quite small, and we thus do not consider the VBF-induced production further. 
Note that the $\beta_L^{32} = 0.2$ and $10^{-3}$ curves are practically indistinguishable, as for couplings smaller than $\sim 0.2$, the production is dominated by the electroweak contributions. For couplings larger than $\beta_L^{32} \sim 0.5$, the $t$-channel production is dominant for all the masses. The luminosities in Eq.~\eqref{eq:lumi} and the cross sections in Fig.~\ref{fig:pairprod_csx} suggest that as long as $m_{\mathrm{LQ}} \leq \sqrt{s}/2$, we have non-zero number of events in this channel regardless of the value of $\beta_{L}^{32}$.

The drop in cross section above the $m_{\mathrm{LQ}} = \sqrt{s}/2$ threshold for the direct $\mu^+\mu^-$ production is apparent. 
It is clear, however, that given the small background expectations outlined above, pair production will be visible for all $m_{\mathrm{LQ}} \lesssim \sqrt{s}/2$, independent of $\beta_L^{ij}$, provided that the branching ratios are not too small. 
If the other $\beta_L^{ij}$ have some nonzero value, they will lead to additional contributions to the $t$-channel diagram in Fig.~\ref{fig:diag_pairprod}, which can slightly increase the cross section, particularly near the $\sqrt{s}/2$ threshold.

\begin{figure}
\centering
\includegraphics[width=0.48\linewidth]{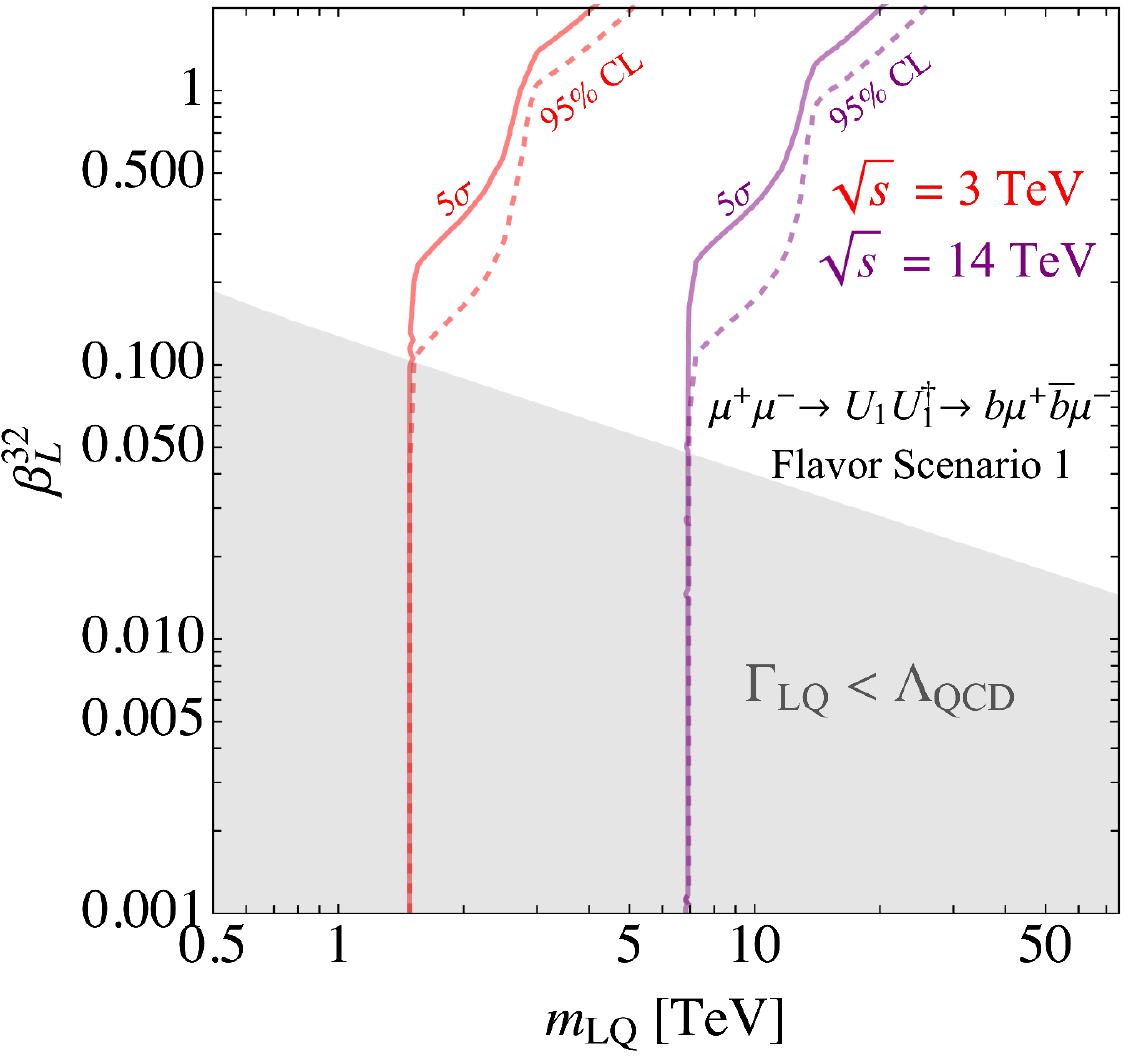}
\hspace{0.1cm}
\includegraphics[width=0.48\linewidth]{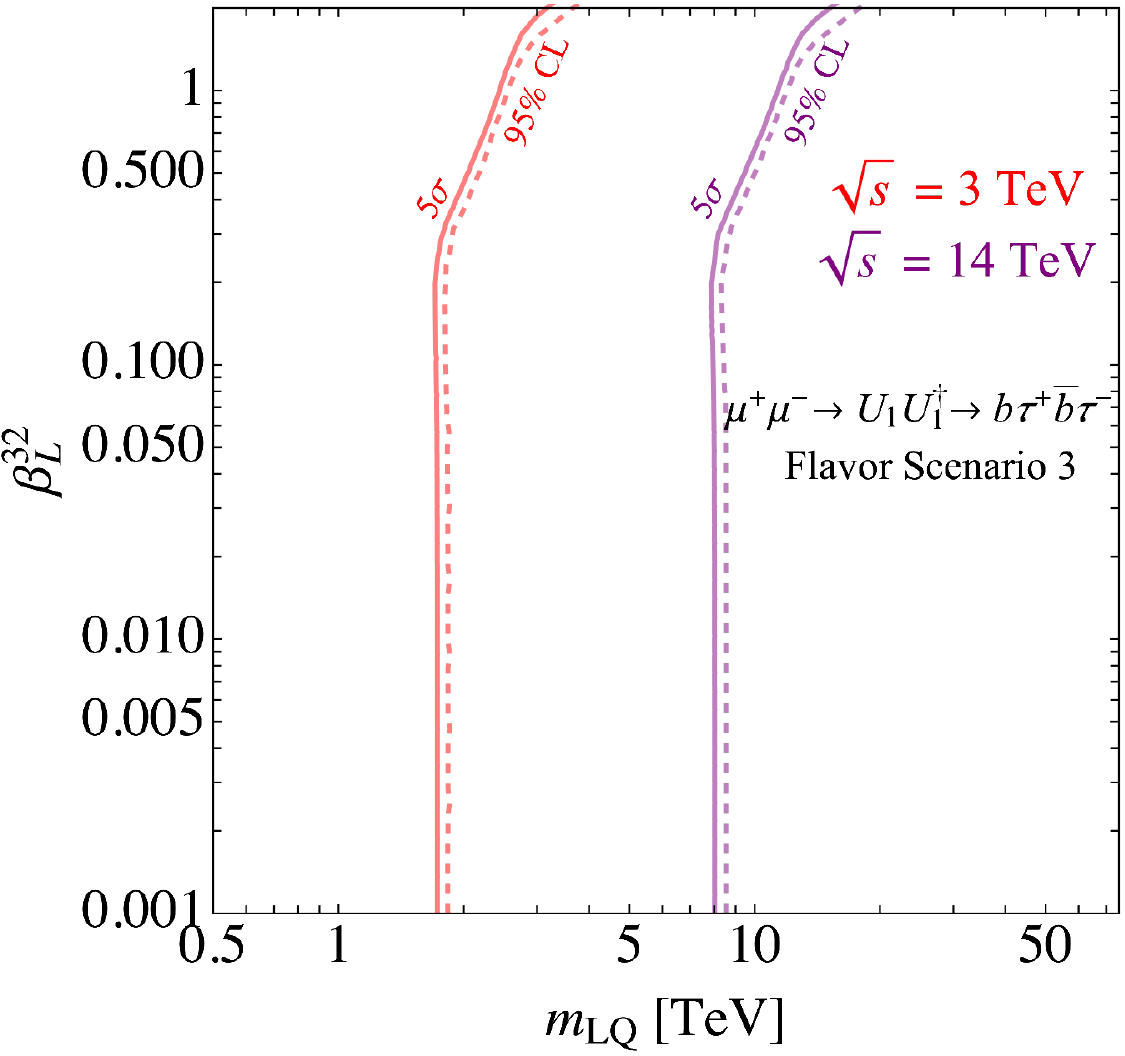}
\caption{Contour plots of the 95\% CL (dashed) and 5$\sigma$ discovery (solid) for pair production of LQ at $\sqrt{s}=3,14$ TeV. We show the reach for flavor scenario 1 (left) and flavor scenario 3 (right). In the gray region, the LQ lifetime is longer than $\Lambda_{\mathrm{QCD}}^{-1}$ and non-perturbative hadronization effects will have to be included for a more accurate result. For sufficiently small couplings the production cross section is dominated by the electroweak production and the bounds become independent of the LQ coupling to the muons.
}\label{fig:vlqpair_contours}
\end{figure}

To estimate the reach of a muon collider, we compute the number of events required to exclude the LQ signal at 95\% C.L. or to claim a $5\sigma$ discovery, based on the background estimates described above.
Details of this treatment are described in detail in App.~\ref{app:stats}.
The resulting constraints for flavor scenarios 1 and 3 are shown in Fig.~\ref{fig:vlqpair_contours}; the bounds on scenarios 2 and 4 are comparable, see Sec.~\ref{sec:comparison}. The solid contours show the $95\%~\textrm{C.L.}$ constraints, while the dashed lines show the $5\sigma$ discovery reach.

For small values of $\beta_L^{32}$, the constraints are essentially constant in mass, as a result of the pure electroweak production of LQ pairs. We see that the electroweak pair production alone will set the reach of a muon collider to roughly $\sqrt{s}/2$. 
For larger values of the muon coupling, the additional $t$-channel production becomes important, and the bounds stretch beyond the $\sqrt{s}/2$ on-shell threshold.
The shaded gray region on the left-hand side shows parameters where the decay width of the LQ is small, and hadronization effects may be important, as discussed in Sec.~\ref{sec:model}.

\subsection{Leptoquark Single Production}
\label{subsec:SP}

\begin{figure}
\centering
\includegraphics[width=0.3\linewidth]{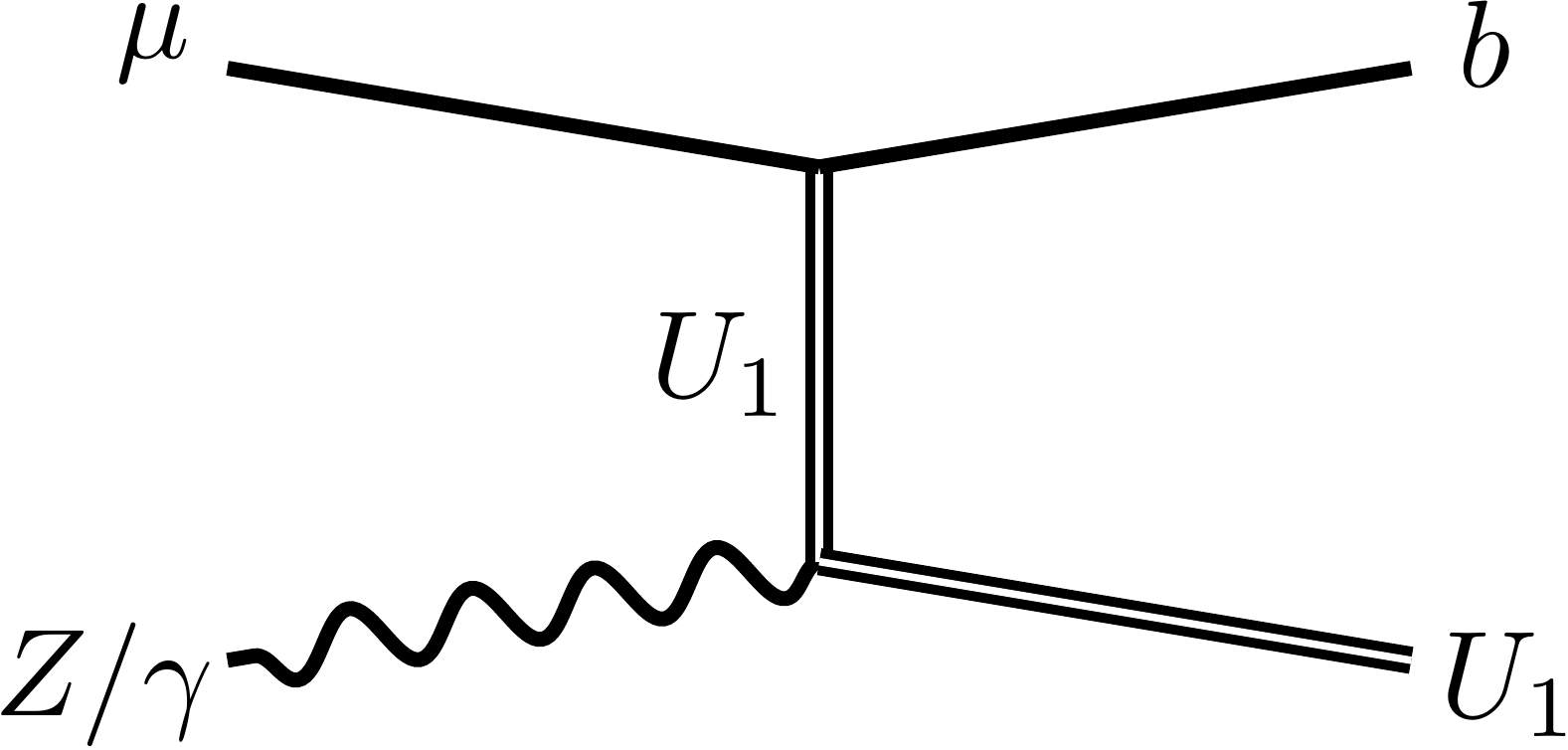}
\hspace{0.25cm}
\includegraphics[width=0.3\linewidth]{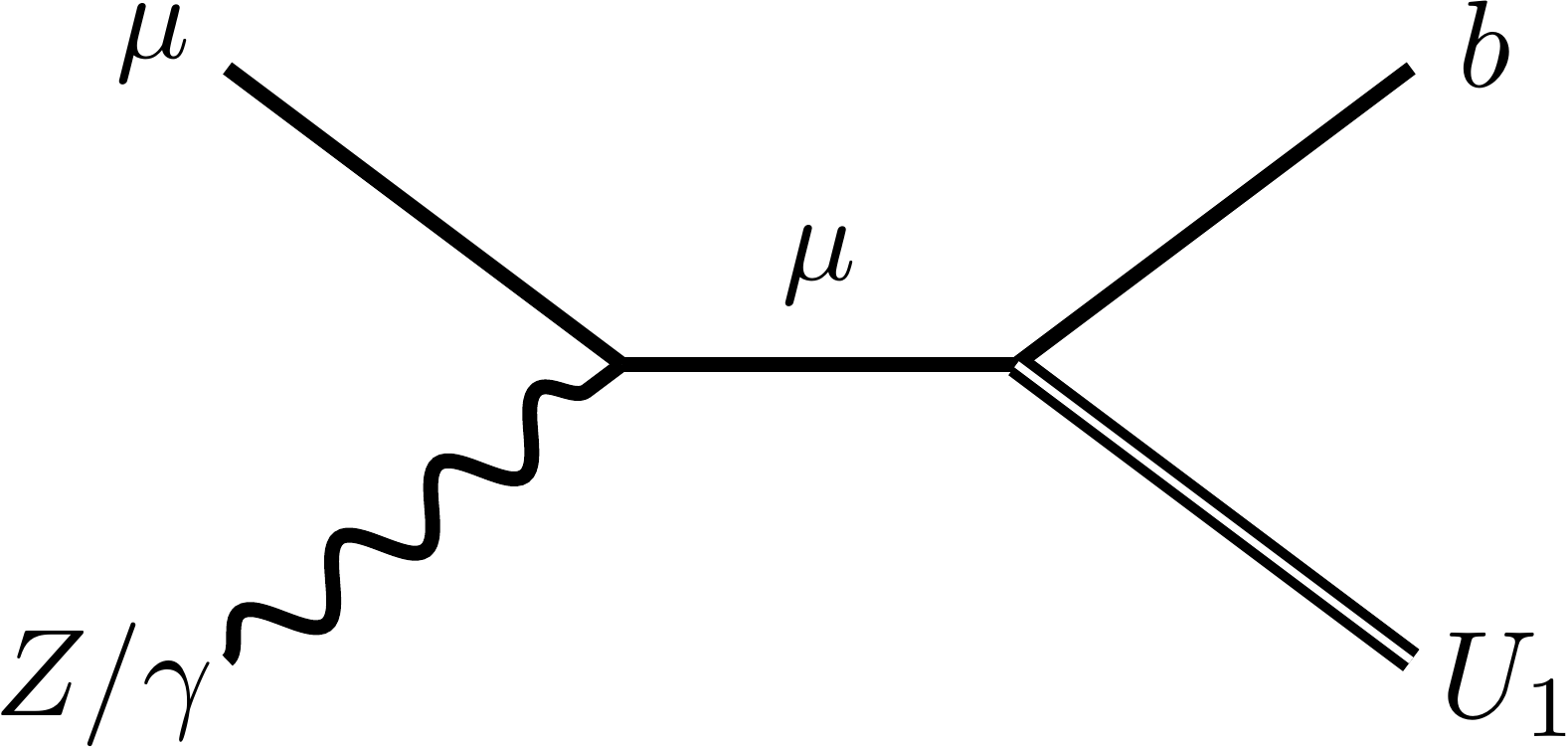}
\hspace{0.25cm}
\includegraphics[width=0.3\linewidth]{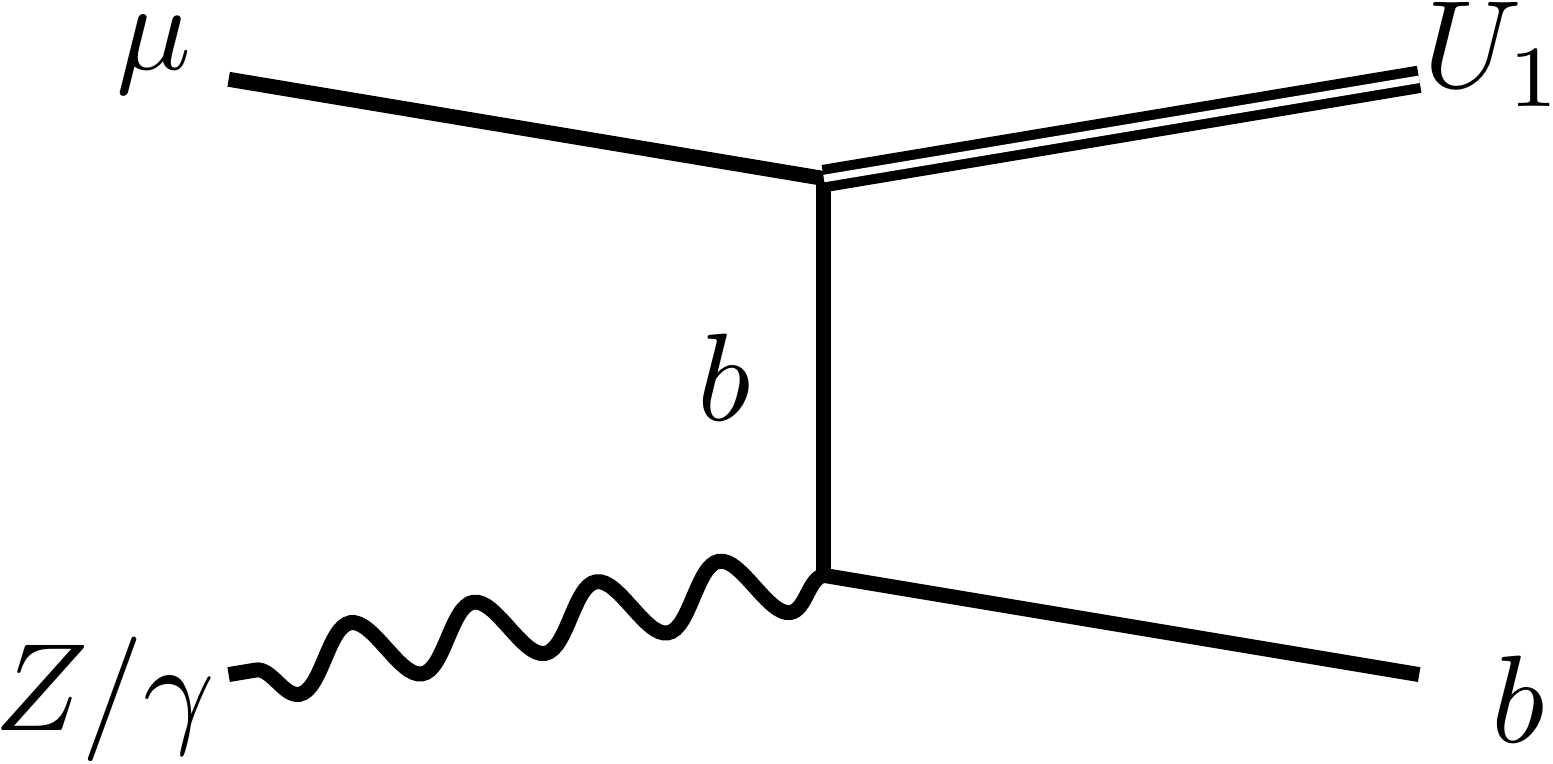}
\caption{
Diagrams leading to single production of LQs. A vector boson from $\mu^+$($\mu^-$) collides with $\mu^-$($\mu^+$) through different channels producing a down-type quark and a LQ.
}\label{fig:diag_singleprod}
\end{figure}

For LQs with masses less than $\sqrt{s}$, single production of LQs can be important. 
We refer to single production as processes $\mu^+ \mu^- \to U_1 d_j + X$, where $X$ is missing energy that escapes down the beampipe or is otherwise not observed. 
The relevant Feynman diagrams for single production of a vector LQ are shown in Fig.~\ref{fig:diag_singleprod}.
In all these diagrams, we show a collision between a muon and a photon or $Z$ boson, where the vector boson is understood to be emitted at a small angle from the incoming muon beam. 
Besides those in Fig.~\ref{fig:diag_singleprod}, there are also two diagrams with intermediate $t$-channel LQ that lead to the same final states. We included those diagrams in the signal cross sections.
In contrast to the ``barking dog'' diagram that was included in the PP topology in the previous section, single production is characterized by events where the muon that radiates the vector boson is deflected by a small angle and continues in the forward direction at high rapidity, outside the coverage of the detector.

A rigorous computation of the signal rate for this inclusive process would make use of the electroweak parton distribution functions (PDFs) for the vector bosons in the muon~\cite{Costantini:2020stv, Han:2020uid, Han:2021kes, Ali:2021xlw}.
In this study, for simplicity, we will content ourselves with working at fixed order and consider only initial state photons, treated as initial states using the effective photon approximation (EPA)~\cite{Frixione:1993yw}. 
Following ref.~\cite{Han:2020uak}, we modify \texttt{MadGraph5} to include photons from muons using the built-in EPA, evaluated at a dynamical scale $Q = \sqrt{\hat{s}}/2$, where $\sqrt{\hat{s}}$ is the partonic center-of-mass energy.
We cross-checked our results using analytic expressions and the electroweak PDFs from refs.~\cite{Han:2020uid, Han:2021kes}, finding reasonable agreement. 
We do not include the contributions from an initial state $Z$ as these are suppressed both due to the $Z$ mass and the electroweak mixing angle. 

 The dominant decay channel of the LQ depends on the scenarios of Tab.~\ref{tab:flavor_structure}. Similar to the PP channel in the previous section, in scenarios 1 and 2 (3 and 4) from Tab.~\ref{tab:flavor_structure} we focus on the LQ decay channel $\bar{\mu} b$ ($\bar{\tau} b$). We also assume a perfect tagging for all the final state particles.

\begin{figure}
\centering
\includegraphics[width=0.48\linewidth]{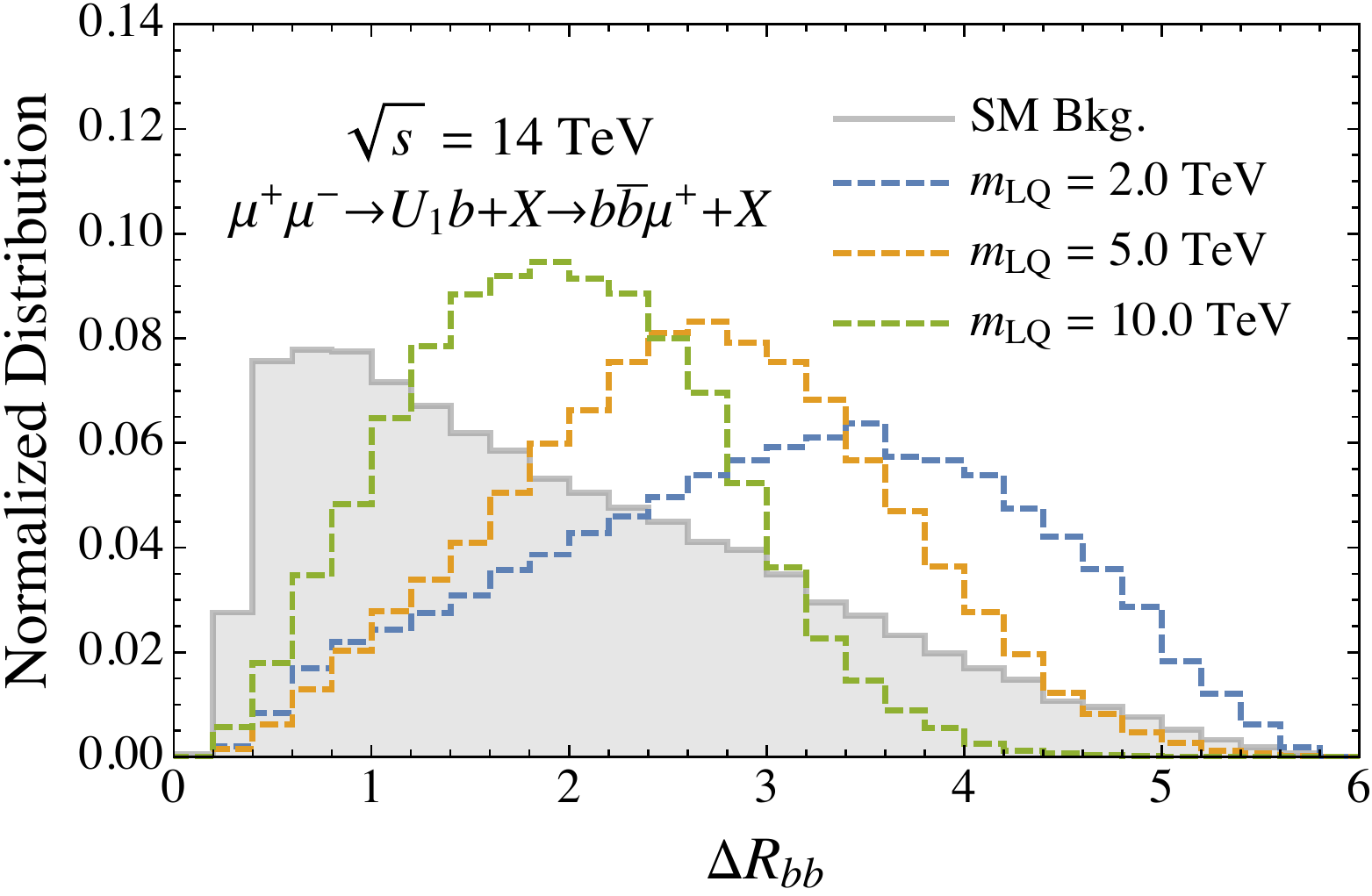}
~
\includegraphics[width=0.48\linewidth]{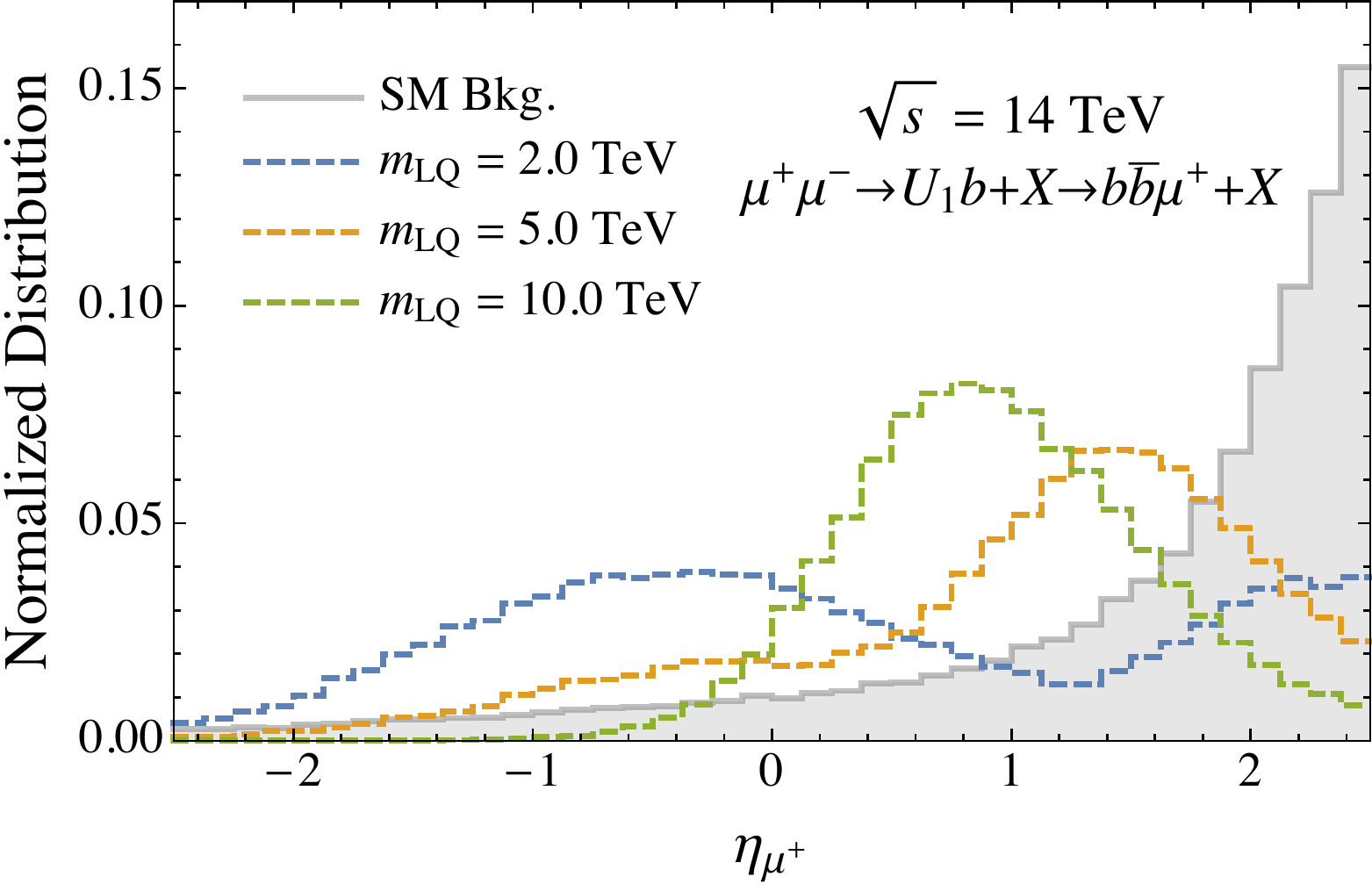}
\caption{Normalized distributions of the angular distance, $\Delta R$ between the $b$-pair (left) and of the pseudorapidity of the $\mu^+$ in single production. The SM background is shown as a gray, shaded histogram while the colored curves show the LQ signal for several values of the LQ mass. The histograms motivate some cuts on $\Delta R_{bb}$ and $\eta_\mu$.}
\label{fig:singleprod_dists}
\end{figure}

The backgrounds to single production in the Standard Model arise from processes very similar to the PP backgrounds, but where one of the final state leptons (muon or tau, depending on the flavor scenario of interest) falls outside the detector coverage, or is otherwise unobserved.
As in the PP background, the SM contributions typically lead to events with a $b\bar{b}$ pair very boosted, and these can be significantly suppressed with a cut $\Delta R_{bb} > 0.5$. 
As shown in Fig.~\ref{fig:singleprod_dists}, the observed lepton tends to be far more central in the LQ signal process than in the SM, where it tends to be forward to balance the unobserved lepton.
We can thus further mitigate the SM background by requiring $|\eta_{\ell}| < 1.5$ for scenarios 1 and 2. 
A similar cut on the visible $\tau$ in scenarios 3 and 4 reduces the expected SM background to $\lesssim 1$ event.

\begin{figure}
\centering
\includegraphics[width=0.48\linewidth]{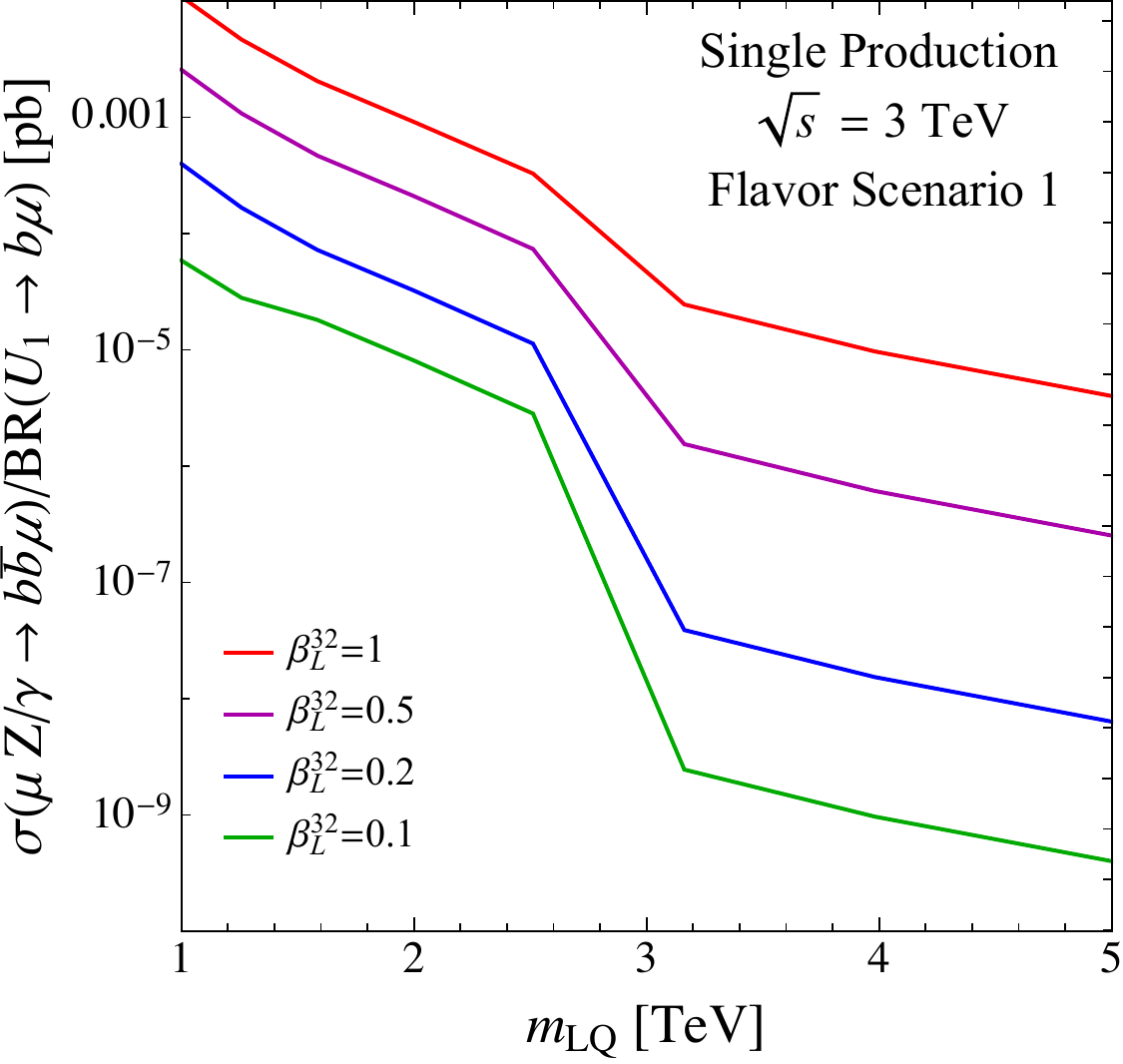}
~
\includegraphics[width=0.48\linewidth]{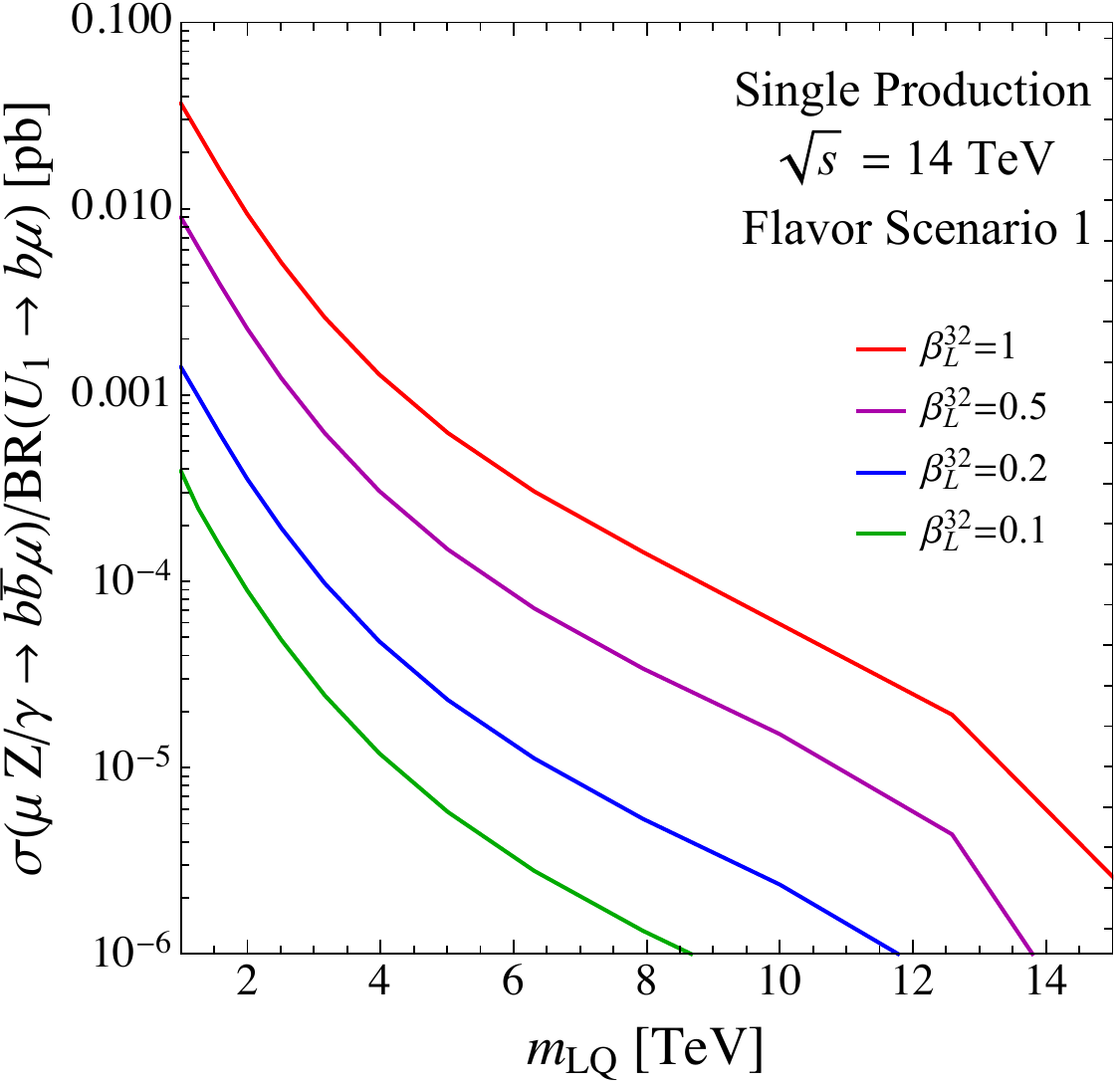}
\caption{SP cross section normalized by the branching ratio of the LQ for the $U_1$ at a COM energy of 3 TeV (left) and 14 TeV (right) for flavor scenario 1. The cross section strongly depends on the coupling to muons and the LQ mass. For $m_{\mathrm{LQ}} \leq \sqrt{s}$, where the LQs can be produced on-shell, the cross section scales like $(\beta^{32}_L)^2$, and it scales as $(\beta^{32}_L)^4$ for higher masses.}
\label{fig:xsec_singleprod}
\end{figure}

Contrary to the PP mode, the single production depends on both the electroweak couplings of the LQ as well as the direct coupling to muons, and vanishes in the limit $\beta_L^{32} \to 0$. 
In Fig.~\ref{fig:xsec_singleprod} we show the LQ SP cross section as a function of $m_{\mathrm{LQ}}$, after the cuts described above, for a few representative values of the coupling at $\sqrt{s} = 3\,\textrm{TeV}$ (left) and $\sqrt{s} = 14\,\textrm{TeV}$ (right). 
As long as the LQ is on-shell, we see that the cross section scales as $(\beta_L^{32})^2$, as expected from the diagrams in Fig.~\ref{fig:diag_singleprod}.

\begin{figure}
\centering
\includegraphics[width=.48\linewidth]{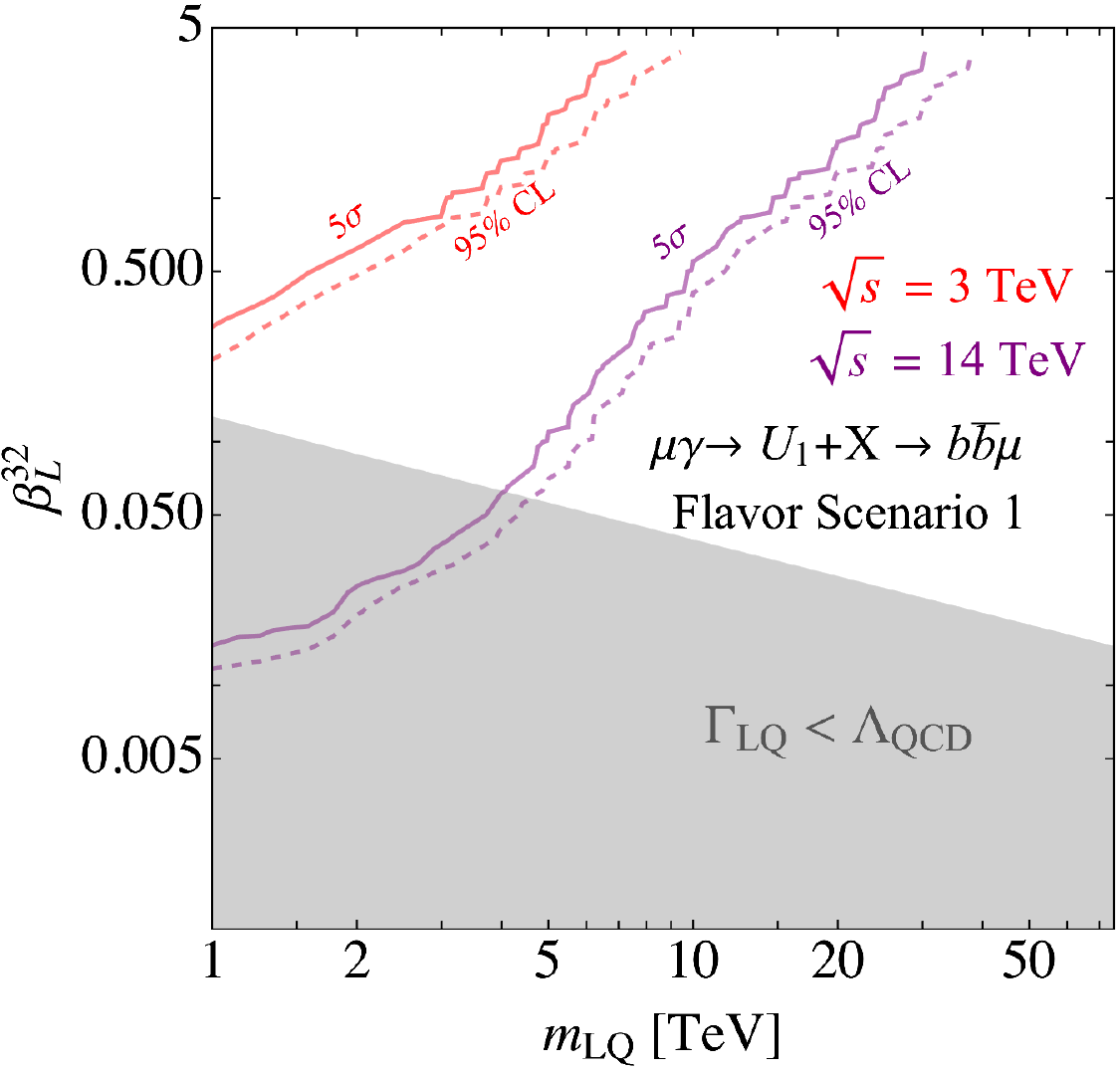}
\hspace{0.1cm}
\includegraphics[width=.48\linewidth]{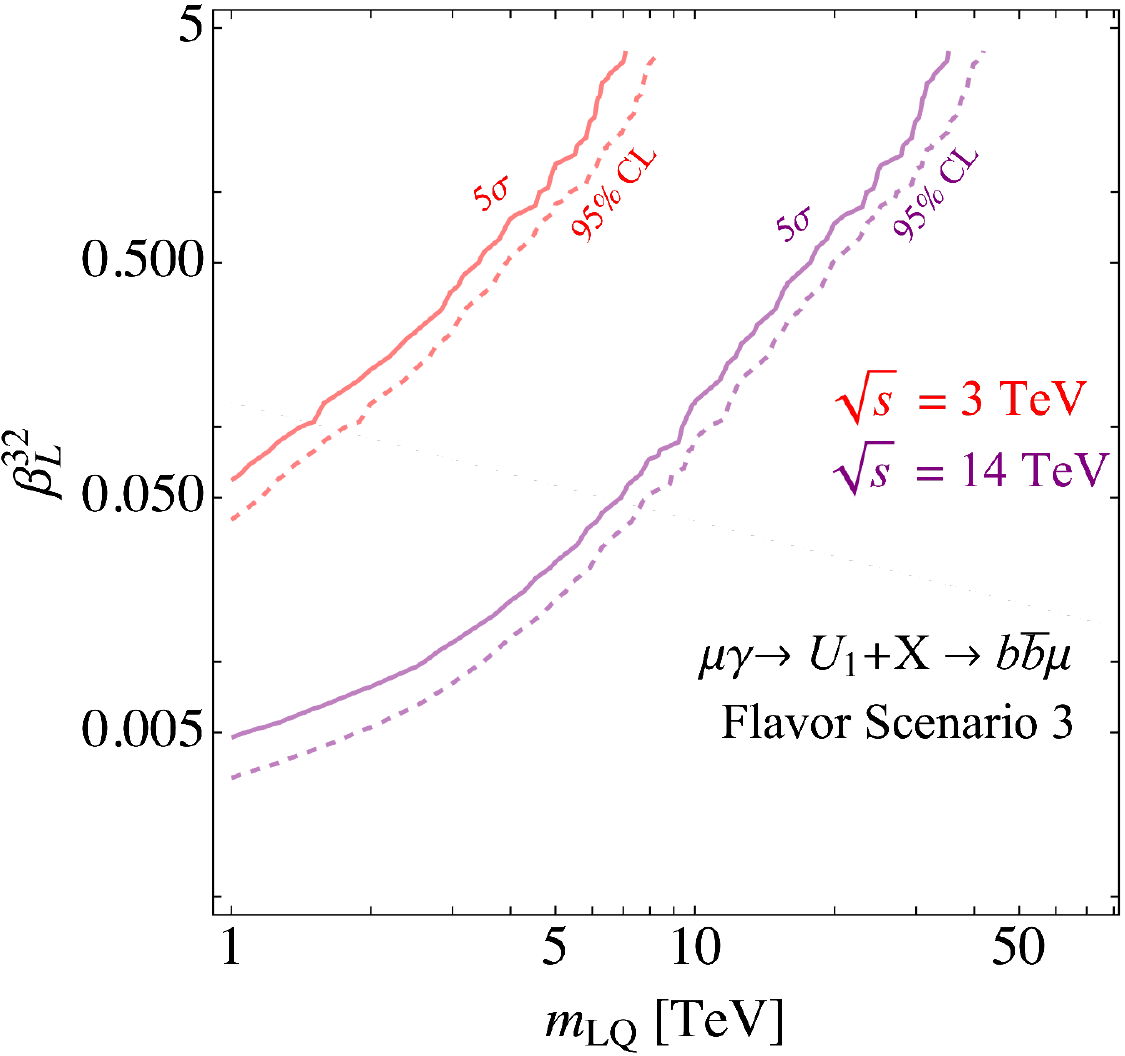}
\caption{Contour plots of the 95\% CL (dashed) and 5$\sigma$ discovery (solid) for single LQ production at $\sqrt{s}=3, 14$ TeV. We show the reach for flavor scenario 1  with one muon in the final state (left), and flavor scenario 3 with a tau in the final state (right). In the gray region, the LQ lifetime is longer than $\Lambda_{\mathrm{QCD}}^{-1}$ and non-perturbative hadronization effects will have to be included for a more accurate result.}
\label{fig:singleprod_contours}
\end{figure}

As is clear from Fig.~\ref{fig:xsec_singleprod}, only large values of the coupling provide a sizeable signal cross section. 
The resulting constraints in the $\beta_L^{32}$ vs. $m_{\textrm{LQ}}$ plane are derived in the same manner as for the PP limits described in the previous subsection, and shown in Fig.~\ref{fig:singleprod_contours}. 
The left panel shows the bounds for flavor scenario 1, while the right panel shows the same constraints for scenario 3.
We see that the shape of the bounds is complementary to the PP bounds, as expected, but they fall off both at high masses and for small muon couplings.
Note however, that the SP constraints gain more power for larger $\sqrt{s}$ than would be expected from a naive scaling, due to the logarithmic enhancement of the photon flux in the EPA.
While the production signal could in principle be improved with a more carefully optimized analysis, we find that the constraints on LQs too heavy to be seen in pair production are weaker than the DY bounds discussed in the next section.

\subsection[Drell-Yan]{Drell-Yan}
\label{subsec:DY}

\begin{figure}
\centering
\includegraphics[width=0.3\linewidth]{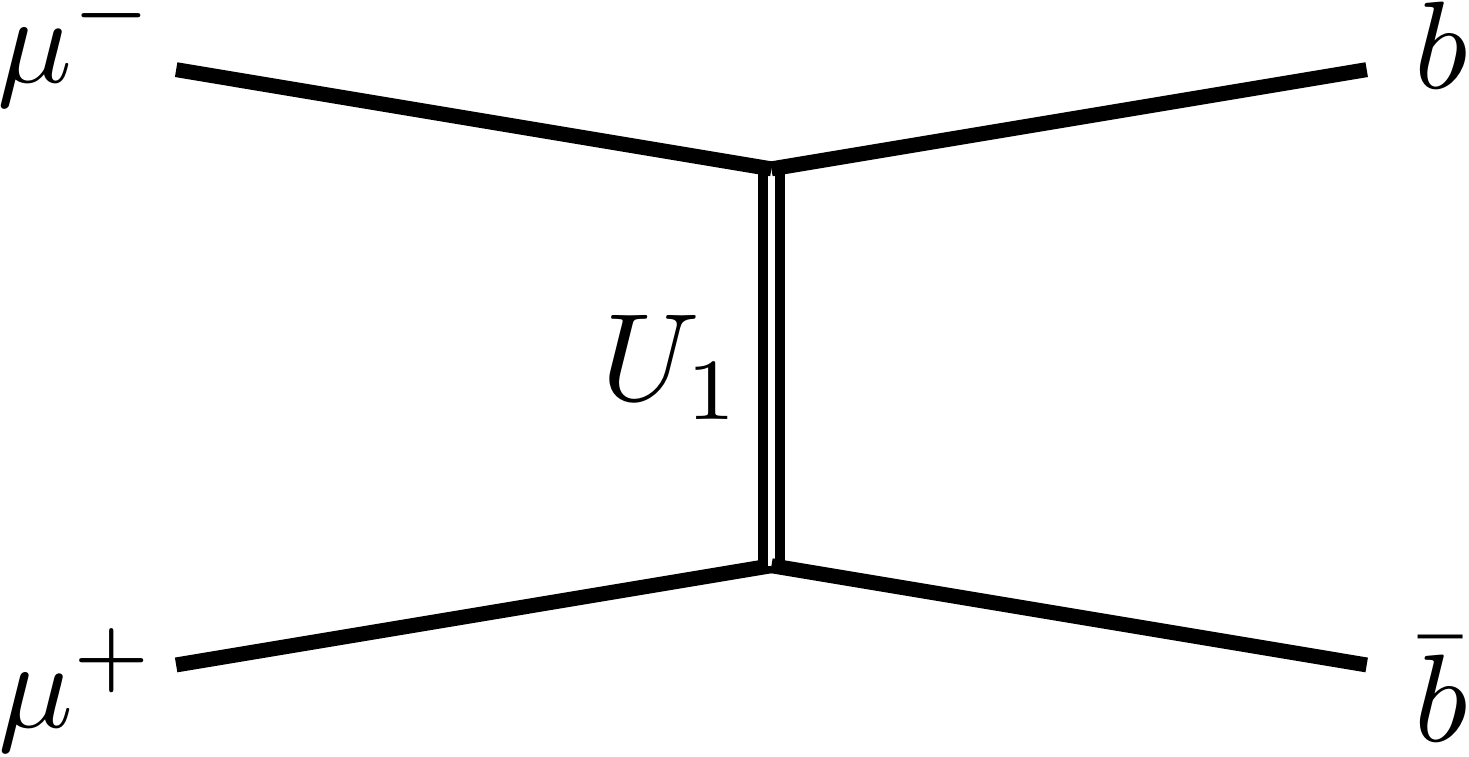}
\qquad
\includegraphics[width=0.3\linewidth]{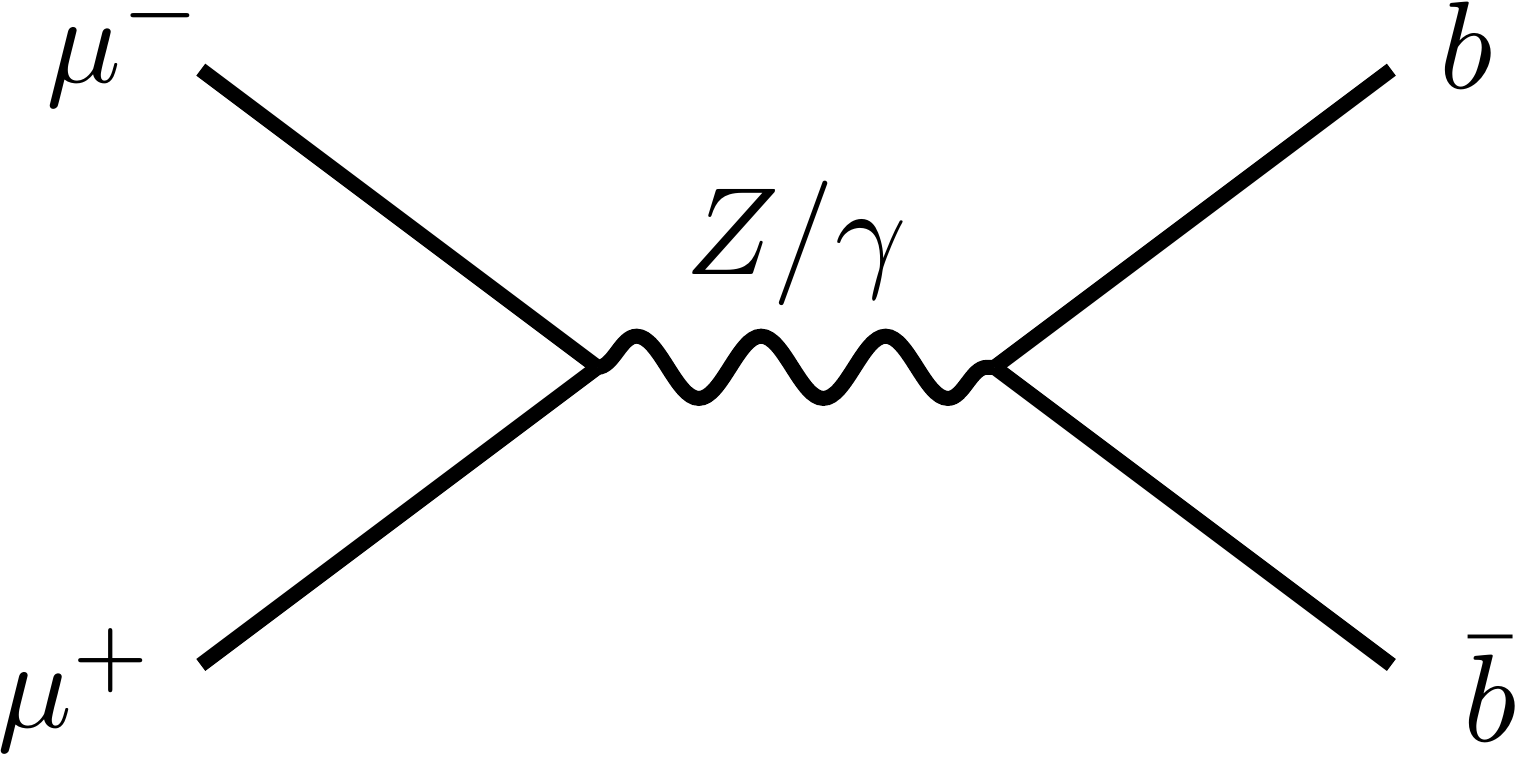}
\caption{
Contribution to DY dijet production from LQ exchange (left) and SM (right). We can use the interference of these two diagrams to look for the LQ signal.
}\label{fig:diag_drellyan}
\end{figure}

Here we consider the LQ interference with the SM DY processes.\footnote{Technically, this is the (Drell-Yan)$^\dagger$ process since we annihilate two leptons into two quarks.}
The parton-level final state comprises two back-to-back quarks. 
The LQ exchange occurs only in the $t$-channel. 
The primary contribution to the cross section in the kinematic regime of interest is the interference of the LQ diagram with the SM $s$-channel DY process. 
The parton-level diagrams are shown in Fig.~\ref{fig:diag_drellyan}.

Unlike the SP and PP channels, the DY process does not contain an $s$-channel LQ. 
Because the effect is in the kinematic distributions and via interference with the SM, there is sensitivity to the NP signal even at higher masses, as the effects scale as $1/m_{\textrm{LQ}}^2$. 
Since the contribution to Drell-Yan via LQ exchange is entirely due to $\beta_L$ and does not depend on the electroweak couplings, the channel loses sensitivity in the small coupling regime. We focus on the $b$ jet final states, which means the signal is sensitive to only the $\beta_L^{32}$ coupling, i.e it is independent of other potential LQ decay channels. 
Furthermore, the DY reach is insensitive to modified gauge interactions, so any constraints apply regardless of $\tilde{\kappa}_U$.

\begin{figure}
    \centering
    \resizebox{\columnwidth}{!}{
    \includegraphics[width=.45\textwidth]{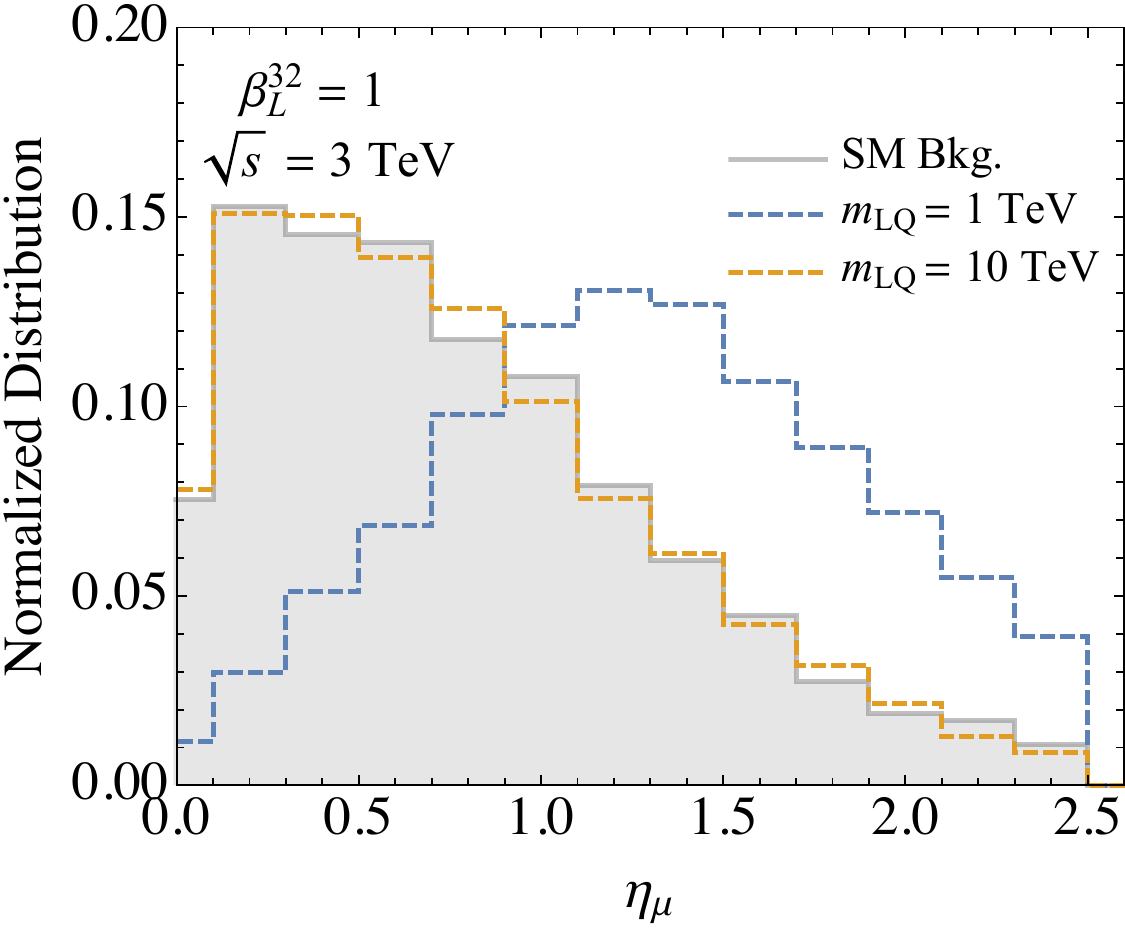} \qquad
    \includegraphics[width=.45\textwidth]{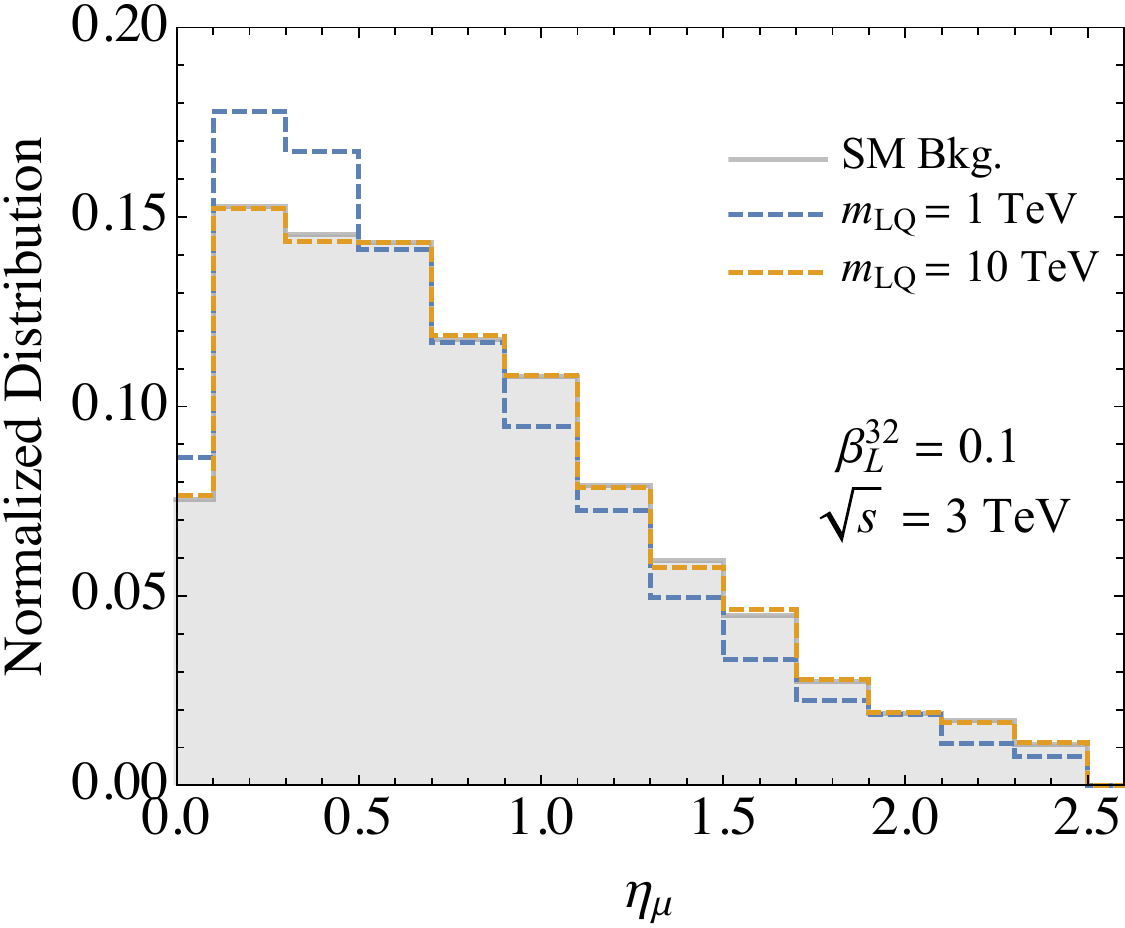}
    }
    \caption{Distribution of DY events in $\eta$ for two different values of LQ mass and its coupling to $\mu_L$ and $b_L$ ($\beta_L^{32}$). We use $\sqrt{s} = 3$~TeV for generating these results.  We observe that SM distribution (gray) can be significantly different from the LQ model prediction, which is a consequence of different SM and LQ diagram topologies. We use these different distributions to search for the LQ signal in this channel. }
    \label{fig:DY_histograms}
\end{figure}

Due to the distinct topology of the LQ contribution to DY production, the presence of LQs will modify the kinematics of jet-pair production, which can be seen in $\eta$, $\theta$, or jet $p_T$ distributions. 
Since this process has a two body final state, these quantities are trivially related and we choose to consider only the $\eta_j$ distribution. 
In Fig.~\ref{fig:DY_histograms}, we show the event distribution in $\eta$ for a few different LQ masses and couplings. 
In the regions of parameter space that the $t$-channel LQ contribution dominates, e.g. low LQ mass or large couplings, the distribution is shifted to larger values of $|\eta|$.

As is clear from Fig.~\ref{fig:DY_histograms}, the overall distribution of events in the presence of the LQ signal can be quite different from the SM.
We leverage the shape-dependence of the distributions in $\eta$ to derive projected $95\%$ C.L. exclusion bounds as well as the $5\sigma$ discovery reach of a \muc{} from this DY channel. 
To do so, we adopt a frequentist approach and use the standard likelihood ratio test statistic to calculate these bounds. 
Further details on our likelihood analysis and calculation of these bounds are included in App.~\ref{app:stats}.

\begin{figure}
    \centering
    \resizebox{0.6 \columnwidth}{!}{
    \includegraphics{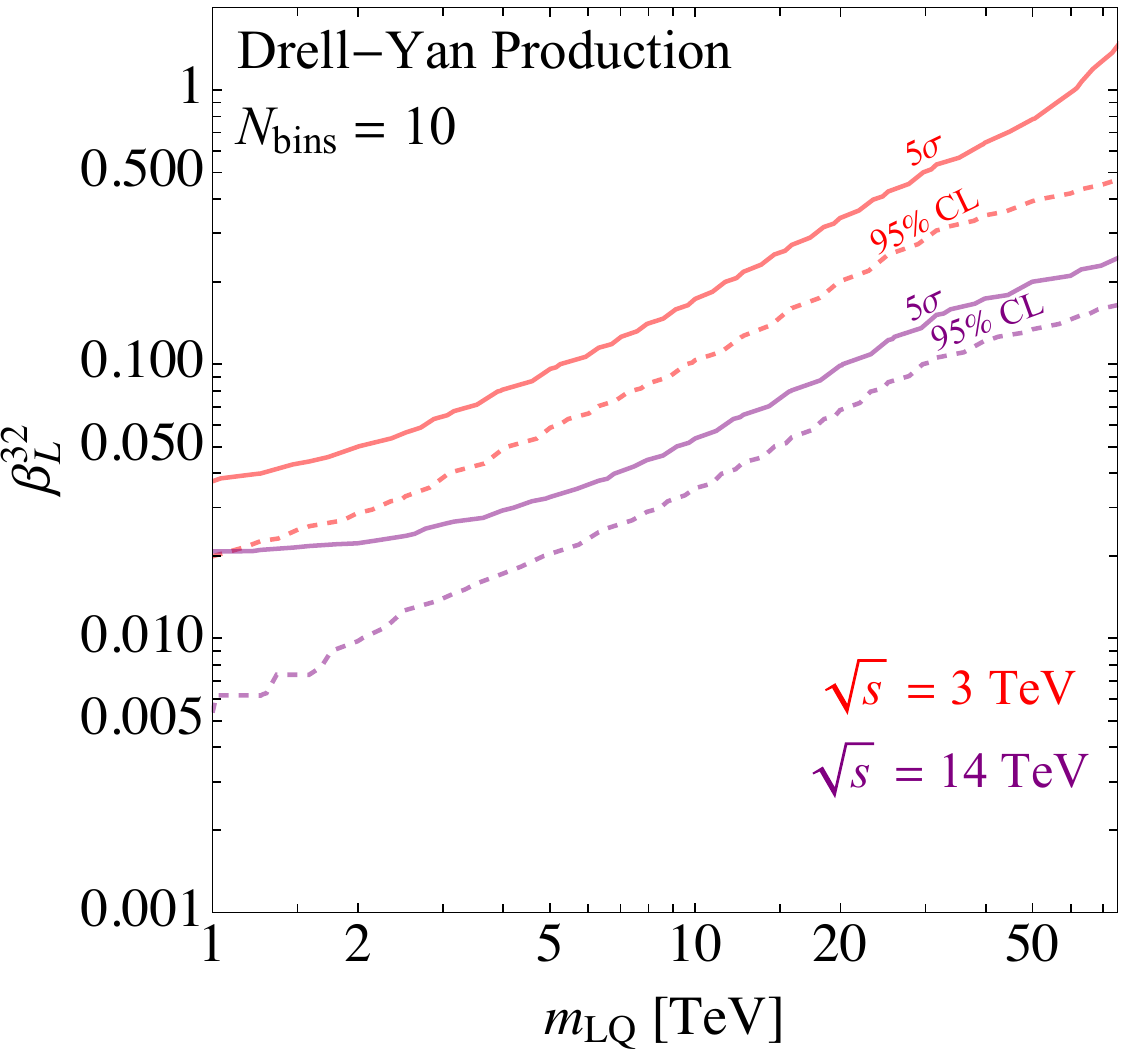}
    }
    \caption{The $95\%$ CL exclusion bound (dashed) and the $5\sigma$ discovery (solid) reach of the DY interference channel with $\sqrt{s} = 3, 14$ TeV. In calculating these bounds we neglected the systematic uncertainties. We also use 10 bins in $\eta$ for the final be jets. The DY channel bounds only depend on the $\beta^{32}$ couplings, thus are the same across the four scenarios of Tab.~\ref{tab:flavor_structure}. }
    \label{fig:DY_result}
\end{figure}

In Fig.~\ref{fig:DY_result} we show the reach of a \muc\ with COM energies $\sqrt{s} = 3$ and $14\,\textrm{TeV}$. 
We compute this reach after binning the events into 10 bins spanning the full detector range in $|\eta|$ ($|\eta| \leq 2.5$). 
We find that increasing the number of bins does not significantly increase the sensitivity. 
Note that in deriving these results, we neglect any systematic uncertainties, which could be easily incorporated into this type of analysis.

While we considered the distribution of events in $\eta$ for this channel, one could equivalently consider the distribution of events in e.g., $\cos\theta$ or $p_T$ due to the one-to-one correspondence with two-body kinematics. 
We confirmed that using any of these variables does not perceptibly change the bounds in the parameter space.

\section{Combination and Comparison to Flavor Constraints}

\label{sec:comparison}

\begin{figure}
    \centering
    \resizebox{ \columnwidth}{!}{
    \includegraphics{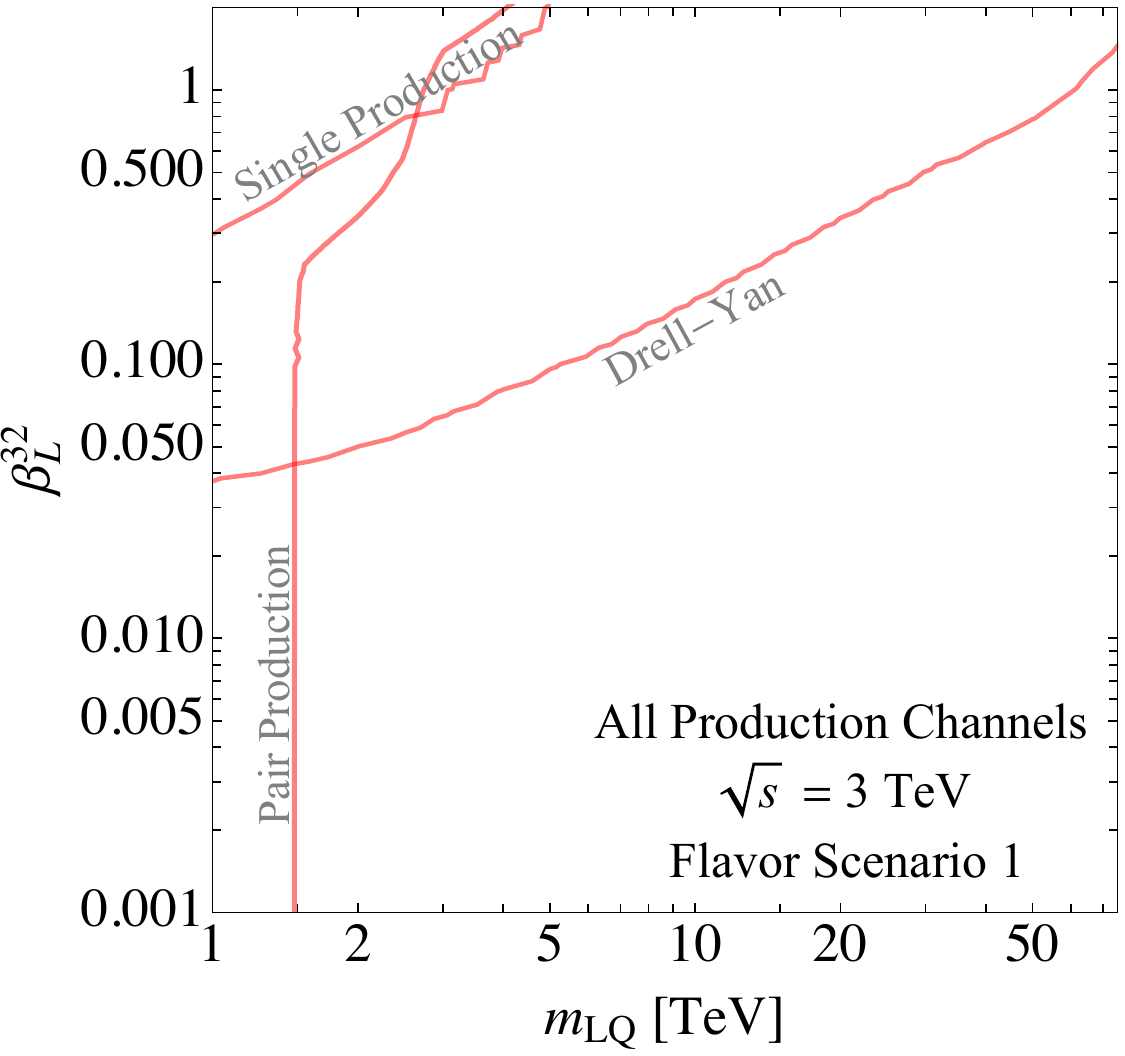}
    \includegraphics{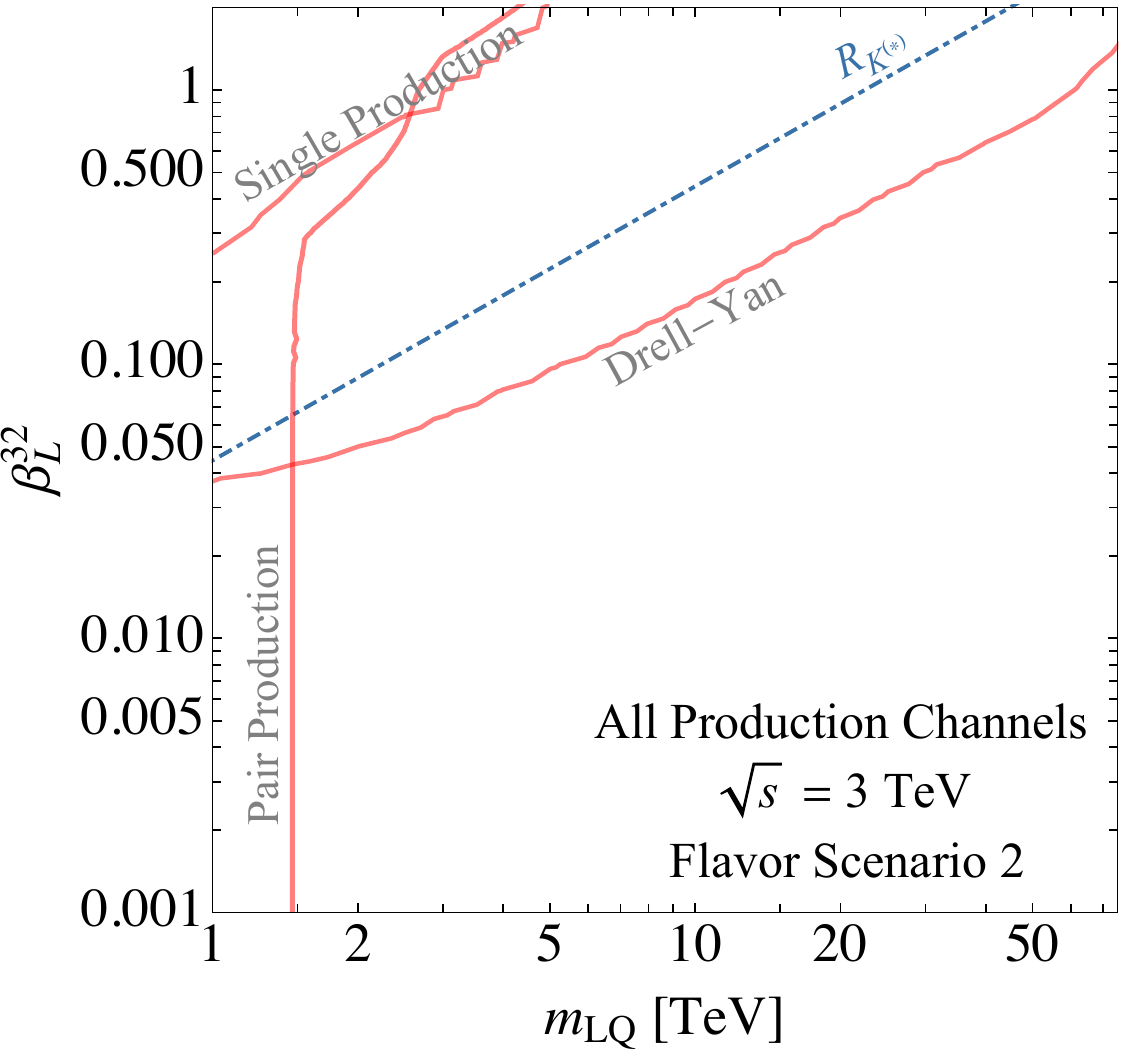}
    }\\
    \resizebox{ \columnwidth}{!}{
    \includegraphics{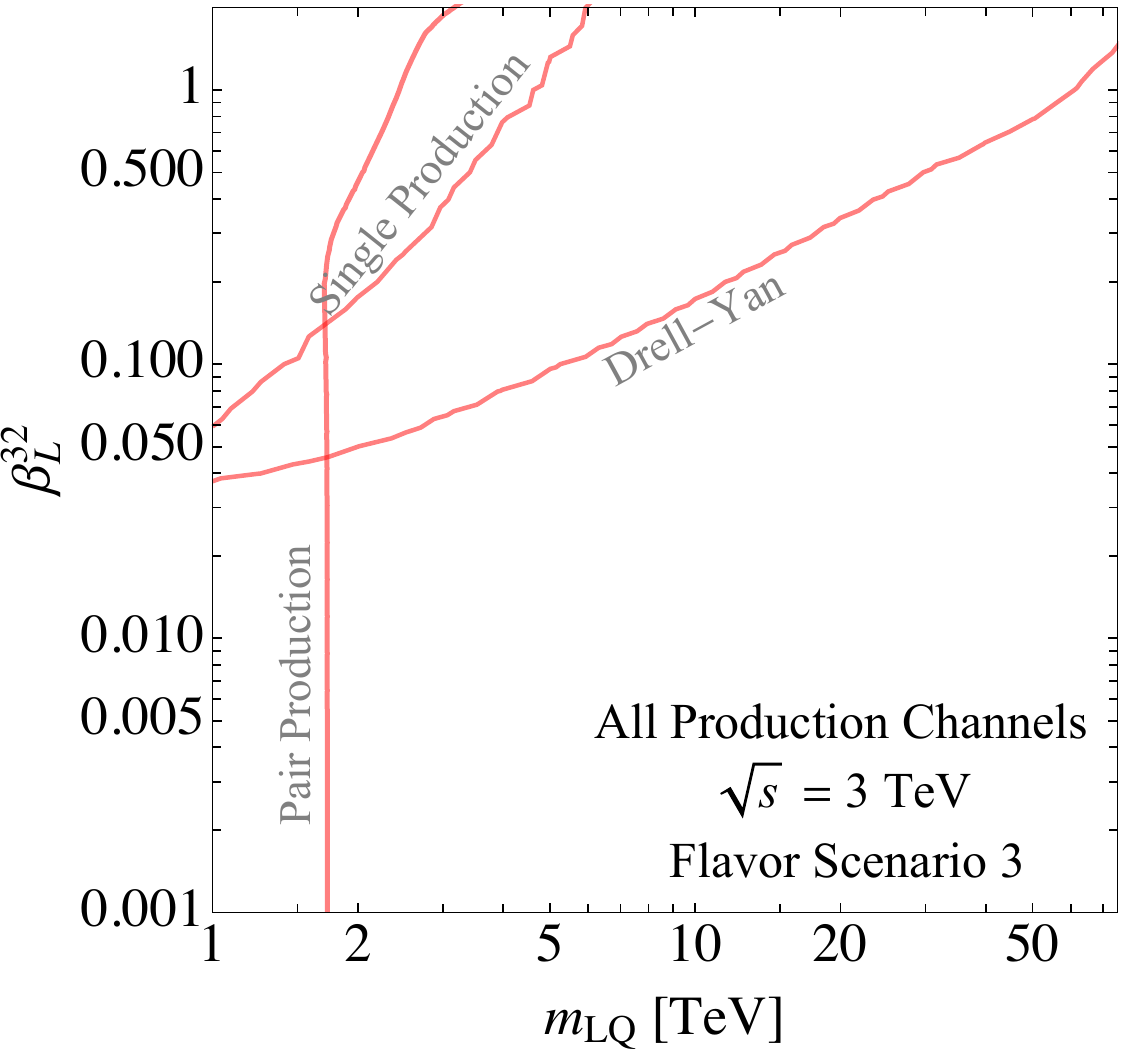}
    \includegraphics{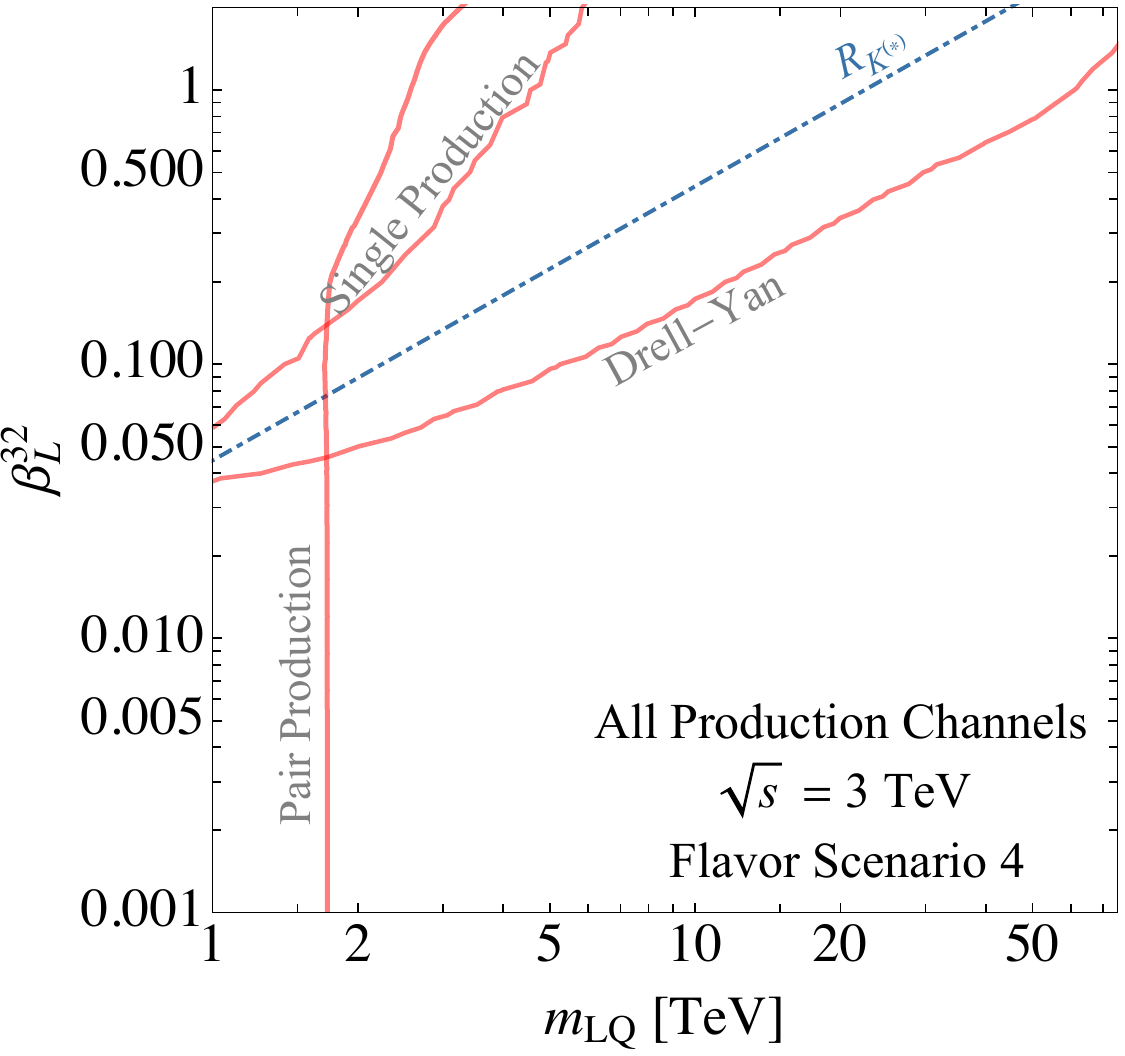}
    }
    \caption{The $5\sigma$ discovery reach of the all channels at $\sqrt{s} = 3$ TeV. Any LQ model in the region to the left or above the red lines can be discovered by the corresponding channel. We show the results for all flavor scenarios presented in Tab.~\ref{tab:flavor_structure}. The final state we search for in the scenarios 1 and 2 (3 and 4) is $\mu b \bar{b}$ ($\tau b \bar{b}$). The DY interference bounds are the same across different scenarios, while the single and pair production can change between the scenarios of top or on the bottom row. Additionally, for flavor scenarios 2 and 4, we include the contours corresponding to the central value of the $R_K$ anomaly. We find that the parameter space explaining this anomaly is completely covered with our proposed searches. The PP channel can cover the low LQ mass of the parameter space, while the DY interference and single production probing the higher masses; the former can probe LQ masses far beyond the intrinsic reach of the collider.}
    \label{fig:combined3}
\end{figure}

\begin{figure}
     \centering
    \resizebox{ \columnwidth}{!}{
    \includegraphics{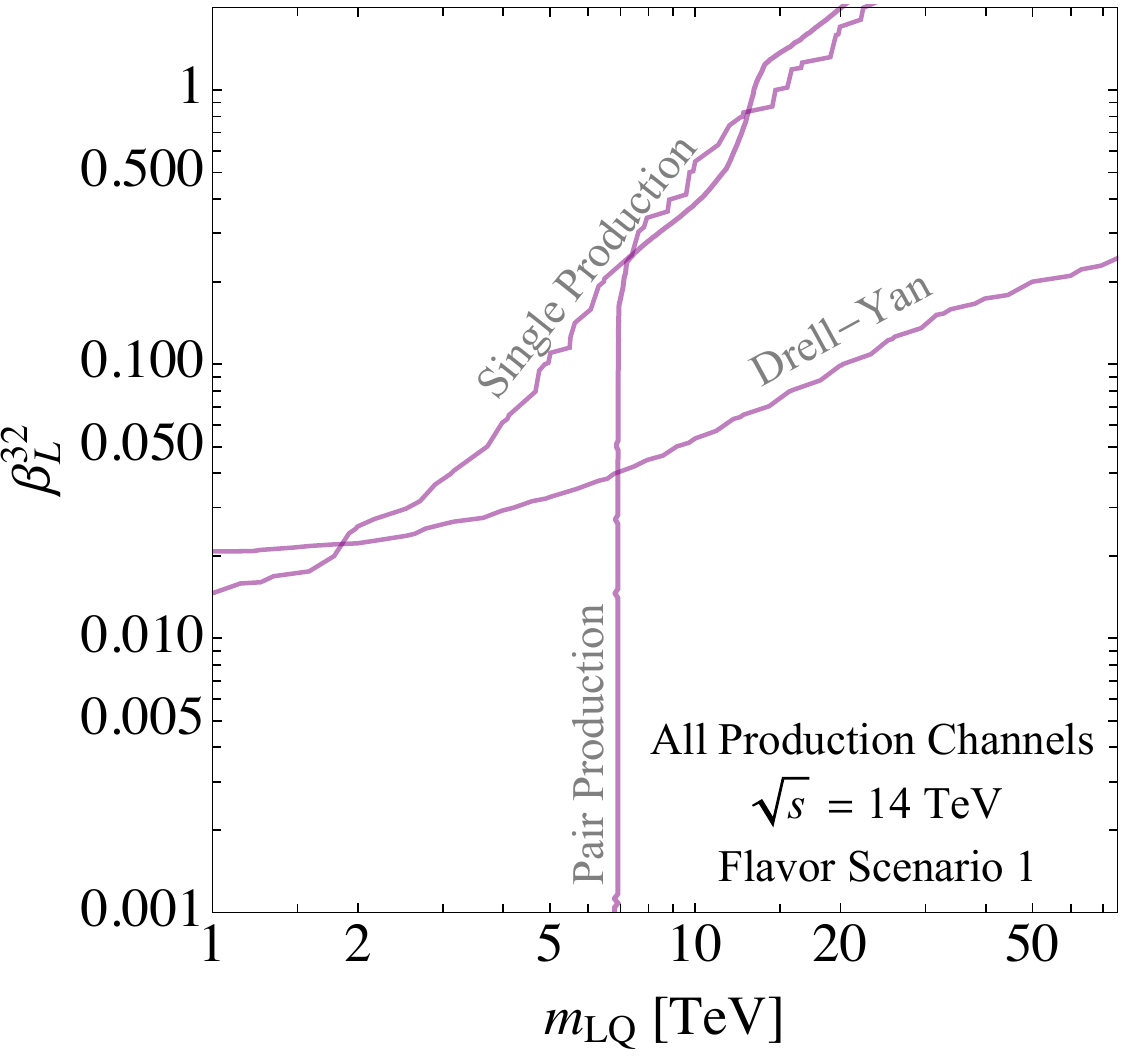}
    \includegraphics{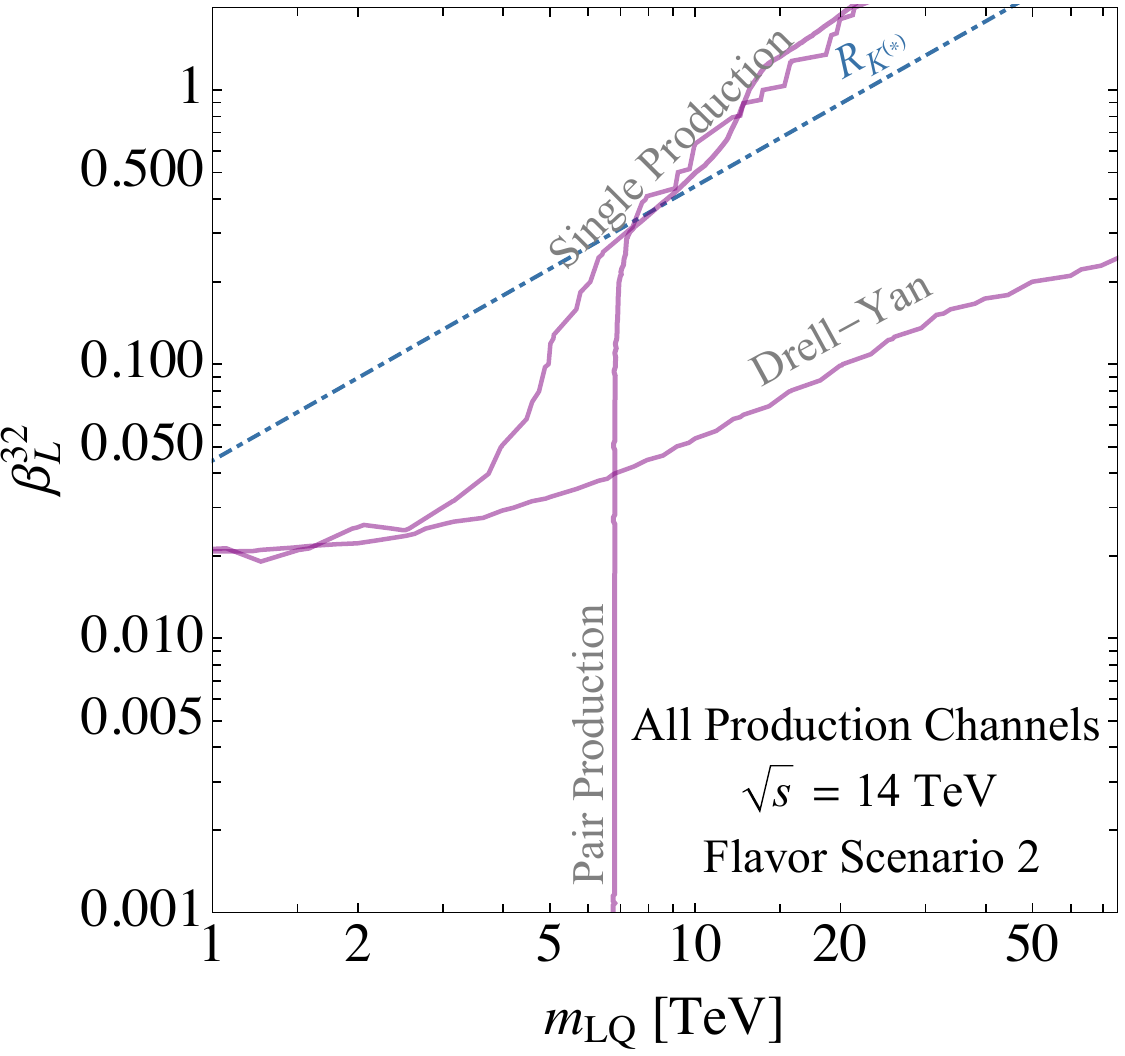}
    }\\
    \resizebox{ \columnwidth}{!}{
    \includegraphics{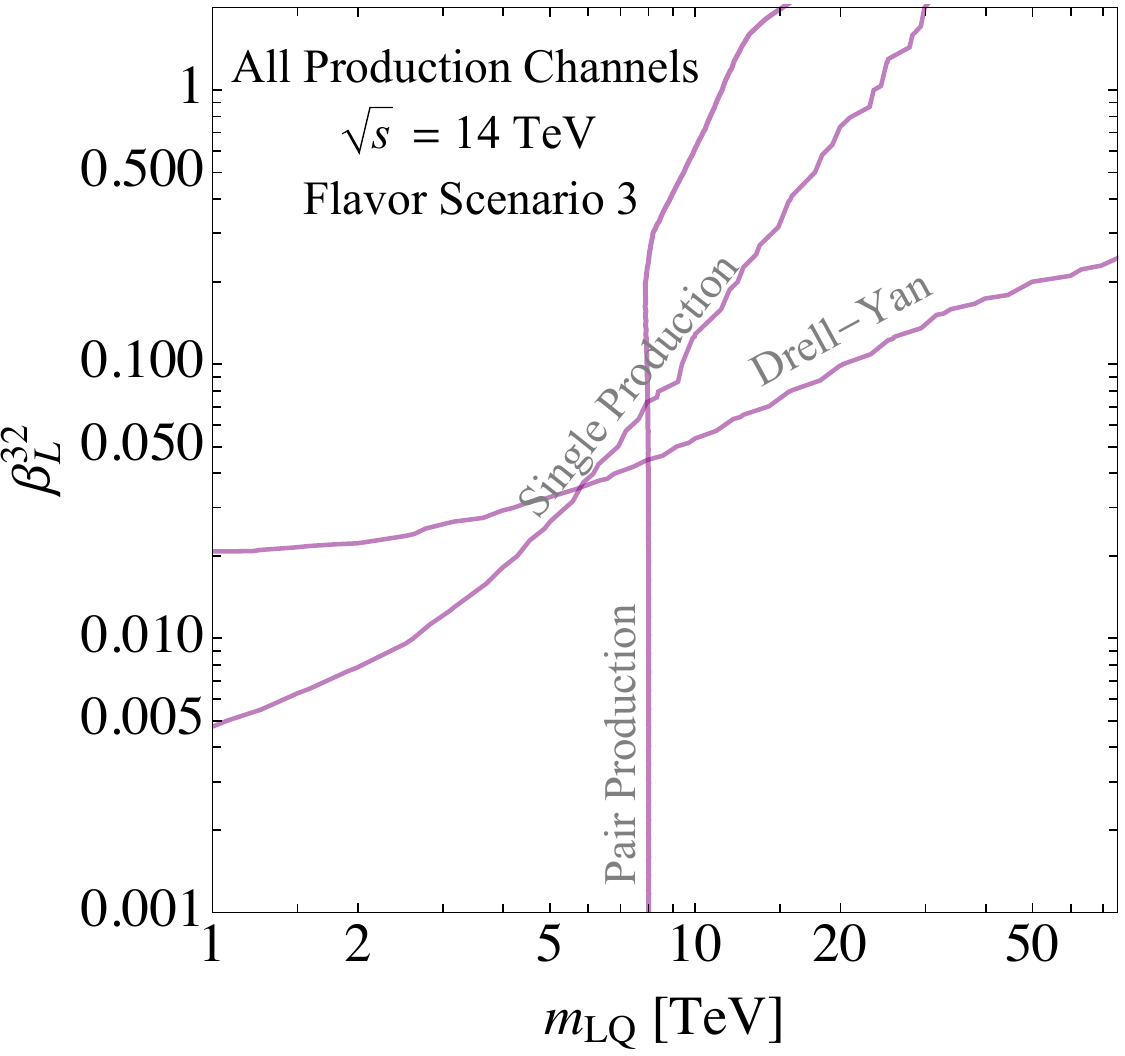}
    \includegraphics{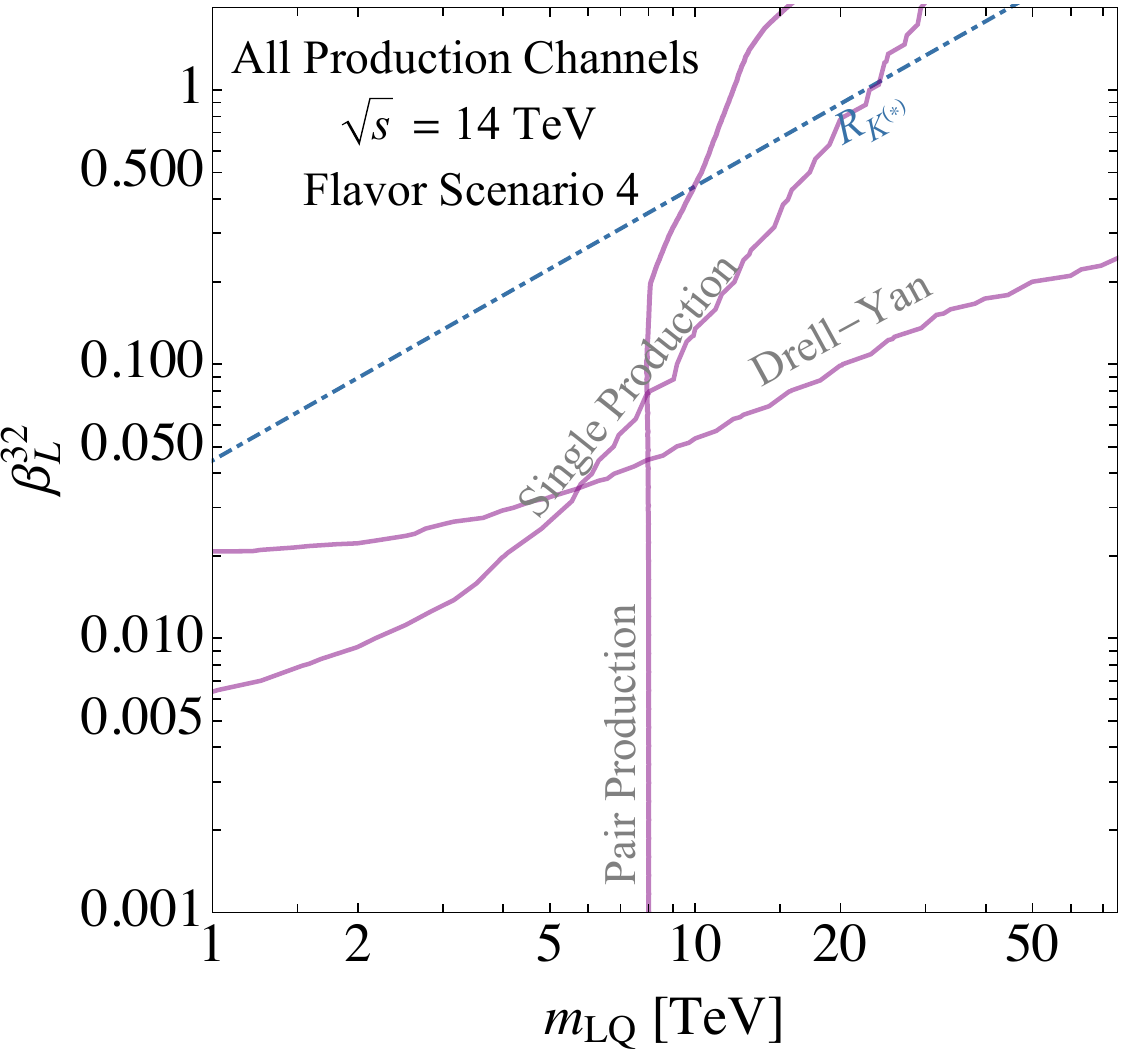}
    }
    \caption{Same as Fig.~\ref{fig:combined3} but for $\sqrt{s} = 14$ TeV. Any LQ model in the region to the left or above the purple lines can be discovered by their corresponding channel.   }
    \label{fig:combined14}
\end{figure}

Let us review the discovery reach of all the channels studied in the previous section. In Figs.~\ref{fig:combined3} and \ref{fig:combined14} we combine the $5\sigma$ discovery bounds from all these channels for COM energies $\sqrt{s}=3$~TeV and $\sqrt{s}=14$~TeV, respectively, and for the four flavor structure scenarios of Tab.~\ref{tab:flavor_structure}. 
As indicated in the previous section, the $95\%$ CL exclusion bounds for each channel is comparable to the discovery reach shown in these figures.

The various LQ production channels have complementary reach in parameter space. As explained in Sec.~\ref{subsec:PP}, the PP contributions only rely on the electroweak gauge couplings, and can probe small Yukawa coupling parts of the parameter space. 
We find that for masses below $\sqrt{s}/2$ the PP channel is the most powerful channel, independent of $\beta_L^{32}$. 
As we go to higher masses, the LQs can not be pair produced on-shell and thus this channel drastically loses its sensitivity. 
For masses above $\sqrt{s}/2$, the SP and DY interference channels can have better discovery reach, provided $\beta_L^{32}$ is not too small.
While the SP channel is not competitive with the combination of Drell-Yan and pair production at $\sqrt{s} = 3\,\textrm{TeV}$, it is noticeably more relevant at $\sqrt{s} = 14\,\textrm{TeV}$ due to the logarithmic enhancement in the photon flux, and might be even more useful for higher center of mass energies that we do not consider here. 
The SP channel becomes weaker for masses approaching $\sqrt{s}$, while the interference with the SM allows the DY mode to bound leptoquarks far beyond the intrinsic reach of the collider for $\beta_L^{32} \gtrsim 0.1$.

The PP bounds on scenarios 3 and 4 (with final state $bb\tau \tau$) is slightly stronger than scenarios 1 and 2 (with final state $bb\mu \mu$) thanks to lower background in their final state. 
In the SP channel, the lower SM background on the final state $\tau b \bar{b}$ in scenarios 3 and 4 makes this channel stronger in these scenarios. The bounds from the DY channel only depend on the $\beta_L^{32}$ coupling and is the same between the four scenarios.

Our results for a muon collider are similar to the analogous classification of the different production modes at the LHC outlined in ref.~\cite{Schmaltz:2018nls}.
There, as above, it was shown that pair production sets the best bounds for small couplings, as long as the LQ is within the mass reach of the collider, while the interference with SM DY production sets the best bound for large couplings.
One qualitative difference is that the DY constraints at a muon collider stretch to much smaller couplings than they do at the LHC. This is in part because we are setting optimistic projections, rather than recasting existing constraints that account for detector effects and systematic uncertainties. However, the DY interference is also intrinsically more sensitive at a high energy lepton collider, as the colliding leptons have a much higher relative partonic luminosity than the necessary partons at a hadron collider.

It is also interesting to compare the reach of a muon collider to future hadron colliders. 
At such a collider, the dominant LQ production mode would be through color production. 
A study of the LQs at the FCC-hh collider was carried out in refs.~\cite{Allanach:2017bta,Allanach:2019zfr}. 
They find a sharp $\sim 10$~TeV bound on the LQ mass from PP diagrams, which agrees with a naive scaling of the current LHC constraints. 
This would slightly outperform even the $14\,\textrm{TeV}$ \muc{} reach in the PP channel for small Yukawa couplings, though a $\sim 20\,\textrm{TeV}$ \muc{} would have comparable reach. 
However, as indicated in Fig.~\ref{fig:combined3} and \ref{fig:combined14}, the SP and DY interference channels can have a better reach in larger masses and coupling values at a \muc. 
We are not aware of a similar study for the reach of FCC-hh in these channels. A study of these channels at FCC-hh is in order before a proper comparison to the \muc~ reach can be made.

\subsection{Flavor Bounds}

With the flavor structure given in Eq.~\eqref{eq:betas}, our model can contribute to various flavor observables, summarized in Tab.~\ref{tab:flavor}. Note that we do not consider a possible contribution to the muon electric dipole moment, $(g-2)_{\mu}$, where there is currently a $4.2\sigma$ discrepancy between the experimental value~\cite{Abi:2021gix} and the Standard Model prediction~\cite{Aoyama:2020ynm}. 
However, to explain the observed anomaly with perturbative couplings, our LQ should be below TeV (see for instance \cite{Kowalska:2018ulj}) in mass, which is already covered by LHC searches \cite{Schmaltz:2018nls}. As a result, for the LQ masses that we are interested in at a \muc{}, the contribution to $(g-2)_{\mu}$ can be neglected.\footnote{Notice that assuming zero $\beta_R^{i2}$ can even further suppress the contribution to $(g-2)_{\mu}$ compared to what is discussed in ref.~\cite{Kowalska:2018ulj}. The contribution of this LQ to $(g-2)_{\mu}$ will still be negligible for masses above a few hundred GeV even if these couplings were included. }

\begin{table}[]
    \renewcommand{\arraystretch}{1.2}
    \centering
    \begin{tabular}{|c|c|c|c|}
    \hline
       Observable & Experimental Bounds & \multicolumn{2}{c|}{Relevant Couplings} \\
       \hline
        $R_{K^{(*)}}$  & $\begin{matrix} R_K = 0.846^{+0.044}_{-0.041} \\ R_{K^*} = 0.685^{+0.113}_{-0.069}\pm 0.047\end{matrix} \quad$ \cite{Aaij:2021vac, Aaij:2017vbb} & \multicolumn{2}{c|}{$\beta_L^{32} \times \beta_L^{22}$} \\
        \hline    
        $\mathrm{BR}\left(B_s \rightarrow \mu \mu \right)$ & $ 3.09^{+0.48}_{-0.44} \times 10^{-9} \quad$ \cite{Bobeth:2013uxa, LHCb:2020zud, Santimaria:2021, Altmannshofer:2021qrr}& \multicolumn{2}{c|}{$\beta_L^{32} \times \beta_L^{22}$} \\
        \hline 
        $R_{D^{(*)}}$  & $\begin{matrix} R_D = 0.340 \pm 0.030 \\ R_{D^*} = 0.295 \pm 0.014\end{matrix} \quad $ \cite{Amhis:2019ckw}  &\multicolumn{2}{c|}{ $\beta_L^{33} \times \beta_L^{23}$ } \\
        \hline
        $R_D^{\mu/e}$  & $  0.995 \pm 0.022 \pm 0.039$ \cite{Glattauer:2015teq} & \multicolumn{2}{c|}{$\beta_L^{32} \times \beta_L^{22}$ } \\
        \hline  
        $\mathrm{BR}\left(\tau \rightarrow \mu\gamma \right)$  & $ < 4.4 \times 10^{-8}\quad$  \cite{Crivellin:2018qmi} &
        \multicolumn{2}{c|}{$\beta_L^{33} \times \beta_L^{32}$} \\
        \hline     
        $\mathrm{BR}\left(\tau \rightarrow \mu\phi \right)$  & $ < 8.4 \times 10^{-8}$ & \multicolumn{2}{c|}{$\beta_L^{23} \times \beta_L^{22}$ } \\
        \hline     
        $\mathrm{BR}\left(D_s \rightarrow \mu\nu\right)$ & $ < 5.49 \times 10^{-3}$ & \multicolumn{2}{c|}{$\beta_L^{22} \times \beta_L^{22}$} \\
        \hline     
        $\mathrm{BR}\left(D_s \rightarrow \tau\nu \right)$ & $ < 5.48 \times 10^{-2}$ & \multicolumn{2}{c|}{$\beta_L^{23} \times \beta_L^{23}$ }\\
        \hline
        $\mathrm{BR}\left( B \rightarrow K \tau \mu \right)$ & $< 2.8 \times 10^{-5}$ & $\beta_L^{32} \times \beta_L^{23}$ & $\beta_L^{33} \times \beta_L^{22}$ \\
        \hline  
        $\mathrm{BR}\left(B_s \rightarrow \tau \mu \right)$ & $ < 4.2 \times 10^{-5}$ & $\beta_L^{32} \times \beta_L^{23}$ & $\beta_L^{33} \times \beta_L^{22}$ \\
        \hline   
        $\mathrm{BR}\left(B_s \rightarrow \tau \tau \right)$ & $ < 2.1\times 10^{-3}$ & \multicolumn{2}{c|}{$\beta_L^{33} \times \beta_L^{23}$ }\\
        \hline  
    \end{tabular}
    \caption{ 
    Various low-energy flavor observables along with the latest experimental result and the combination(s) of couplings relevant for each process at tree-level in the $U_1$ leptoquark model, based on the results of refs.~\cite{Feruglio:2017rjo, Angelescu:2018tyl}. 
    Experimental values are taken from ref.~\cite{Zyla:2020zbs}, unless otherwise noted.
    }\label{tab:flavor}
\end{table}

\begin{figure}
\centering
\includegraphics[width=0.3\linewidth]{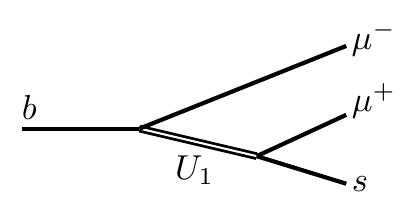}\qquad
\includegraphics[width=0.3\linewidth]{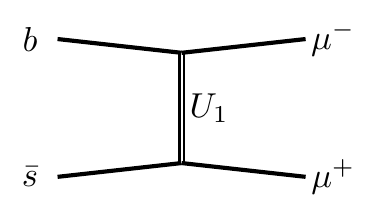}
\caption{Diagrams for the tree-level LQ contributions to the $R_K^{(*)}$ anomaly (left) and $B_s \rightarrow \mu^+\mu^-$ (right). The two contributions are related via crossing symmetry.}
\label{fig:flavor_diag}
\end{figure}

Two of the most sensitive tests of lepton flavor universality (LFU) are the $R_K$ and $R_K^*$ ratios --- the relative rate of $B$ meson decays to final states involving muons over electrons.
Recent experimental measurements of these ratios are $3.1$ and $2.5\sigma$ below the SM prediction, respectively~\cite{Aaij:2021vac,Aaij:2017vbb}.\footnote{See also ref.~\cite{Abdesselam:2019wac} for a measurement of $R_{K^*}$ at Belle with far larger error bars.} 
Leptoquarks can modify these decay rates at tree level, as shown in Fig.~\ref{fig:flavor_diag}. 
Recent theoretical fits to the effective Lagrangian describing these LFU violating decays imply that a LQ with nonzero values of $\beta_L^{22}$ and $\beta_L^{32}$ can fit the discrepancies in data for~\cite{Altmannshofer:2021qrr}:
\begin{equation}
     \frac{\beta_L^{22}\beta_L^{32}}{m_\text{LQ}^2} = 1.98 \times 10^{-3} \text{ TeV}^{-2}
     \label{eq:RKbound}
\end{equation}
These couplings are nonzero in our flavor structures 2 and 4 in Tab.~\ref{tab:flavor_structure}, where we assume that $\beta_L^{22} = \beta_L^{32}$.
The central value in Eq.~\eqref{eq:RKbound} is indicated by a dashed blue line in the right panels of Figs.~\ref{fig:combined3} and \ref{fig:combined14}. We see that even a $3\,\textrm{TeV}$ muon collider has the potential to discover LQs in the parameter space of interest to the flavor anomalies.

A similarly sensitive observable is the leptonic decay $B_s \to \mu^+ \mu^-$, which is related to the process in $R_{K^{(*)}}$ through crossing-symmetry, see Fig.~\ref{fig:flavor_diag}. 
The current measurement of the branching ratio for this decay channel is~\cite{Santimaria:2021}
\begin{equation}
    \text{BR}(B_s \rightarrow \mu^+\mu^-) = (3.09^{+ 0.48}_{-0.44}) \times 10^{-9},
    \label{eq:Bs_new_bound}
\end{equation}
which is compatible with, though slightly smaller than the SM prediction. 
The coupling values that explain the observed $R_{K^{(*)}}$ anomaly also predict this branching ratio will stay within the $1\sigma$ range of the observed value~\cite{Santimaria:2021} and in fact have even better agreement with the observed value~\cite{Altmannshofer:2021qrr}.

In Tab.~\ref{tab:flavor} we show many other observables that can get contribution in our model.
We also show the coupling dependence of each observable; clearly, each observable has contributions only s subset of scenarios that we study. We calculated the bounds from all these observables (using the formulas from refs.~\cite{Feruglio:2017rjo, Angelescu:2018tyl}) and found that none of them are as sensitive as $R_{K^{(*)}}$.

Additionally, electroweak boson decays can be modified by diagrams involving loops with LQs. These effects can be computed with the effective field theory operators described in ref.~\cite{Feruglio:2017rjo, Angelescu:2018tyl}, but provide much weaker constraints than nearly all of the flavor observables.

\section{Conclusions}
\label{sec:conclusion}

A muon collider can open up novel directions in high energy community's quest for physics beyond the SM. 
Recent progress in the experimental front has motivated a closer look into its potential in probing various BSM directions. 
In this work, we studied the reach of such a collider in search for vector LQs, an exotic extension of the SM with unique signals at high energy colliders. 

We focused on a particular LQ called $U_1$, which has drawn a lot of attention over the last few years thanks to its ability to explain various flavor anomalies. 
We considered four representative scenarios for the flavor structure of the LQ couplings to the SM fermions. These ranged from scenarios uniquely suited to production at muon colliders to those most relevant to recent anomalies in $B$-meson decays, which are motivated by UV constructions. 
We explored the differences in the resulting phenomenology at a muon collider, and our analysis can straightforwardly be repeated for any other flavor structure.

We studied three broad classes of signals from this new particle: (i) its pair production, (ii) single production, and (iii) interference with the SM Drell-Yan process. We studied each channel in detail and proposed simple strategies to set limits on the parameter space in the presence of the SM backgrounds.

We found that, thanks to its electroweak charges, a LQ of mass $\sqrt{s}/2$ or lower can be detected through its PP signal. 
This channel eventually loses its sensitivity as we go to larger LQ masses. 
On the other hand, the DY interference signal, although weak at very small couplings to the SM fermions, can potentially probe LQ masses even an order of magnitude heavier than the colliders COM energy with perturbative couplings. 
The SP channel can have a very great reach, even though it is always sub-dominant to the other two channels in the flavor structures that we considered. 
The reach of both SP and PP channels depends on the flavor structure of the LQ coupling to all SM fermions and its branching ratios, while the DY interference channel is only sensitive to a single LQ-SM fermions couplings and can be used to put irreducible bounds on individual couplings.

The LQs are a color charged particle and are expected to be color produced in large numbers at hadron machines. 
Nonetheless, the fact that all of the incoming particles energy can be accessible to a single event and clean environment gives lepton machines an edge over comparable hadron collider options. 
In particular, we found that at a $14\,\textrm{TeV}$ muon collider, DY interference channel is sensitive to LQ masses up to $100\,\textrm{TeV}$ for perturbative couplings, which is completely out of reach for direct production even at a $100\,\textrm{TeV}$ hadron machine.

The high energy muon collider can probe parts of the LQ parameter space beyond the reach of flavor experiments. We found that when the couplings of the LQ to the SM fermions allow for explaining the observed $R_{K^{(*)}}$, and when different relevant couplings are comparable in size, the relevant parameter space of the model is completely probed even at a 3 TeV muon collider.

Our results point to several clear directions for future study. 
We only focused on a specific well-motivated LQ with specific flavor structures. 
Similar analysis should be repeated for other LQs or broader classes of flavor structures. 
Furthermore, in our calculation of the SP signal we used the effective photon approximation and considered only a photon in the initial state. 
While the EPA is a good leading approximation, a more rigorous treatment of the muon PDFs --- particularly for higher energy muon colliders --- is in order. 
Finally, in this work we neglected any detector effects and various sources of systematic uncertainties, so our results are an optimistic projection of the reach of muon colliders in the parameter space of our LQ model.  
However, given that the most optimistic projections would have a muon collider gathering data in no less than 15-20 years, a number of improvements in detector technology can be expected, making these target projections somewhat realistic.

In conclusion, leptoquarks are an exotic class of new physics particle, with interesting collider phenomenology. Our results demonstrate that a muon collider would have unique capabilities for constraining these particles, and perhaps even discovering new physics.

\acknowledgments

We are grateful to David Curtin, Javier Fuentes, Seth Koren, Patrick Meade, Johannes K. L. Michel, Matthew Reece, David Shih, and Xing Wang for helpful conversations. We thank Matthew Buckley, Anthony DiFranzo, Angelo Monteux, and David Shih for their contributions in developing the code-base used in this paper. The work of PA was supported by the DOE Grant Number DE-SC0012567 and the MIT Department of Physics. 
The work of RC was supported in part by the Perimeter Institute for Theoretical Physics (PI), by the Canada Research Chair program, and by a Discovery Grant from the Natural Sciences and Engineering Research Council of Canada. Research at PI is supported in part by the Government of Canada through the Department of Innovation, Science and Economic Development Canada and by the Province of Ontario through the Ministry of Colleges and Universities.
The work of CC and SH is supported in part by the DOE Grant DE-SC0013607. CC is supported in part by an NSF Graduate Research Fellowship Grant DGE1745303. SH is supported in part by the Alfred P. Sloan Foundation Grant No. G-2019-12504.

\appendix

\section{The Statistics}
\label{app:stats}

In this appendix we provide more details on the statistical treatment used to project limits from the various production channels.

For the DY constraints, we use a binned likelihood analysis to find the regions of the $m_{\textrm{LQ}}$-$\beta_L^{23}$ parameter space that are excluded at $> 95\%$ confidence level, assuming the observed events mimic the distribution predicted by the Standard Model. 
We will assume that we have $N_{\mathrm{bin}}$ bins in the forward direction ($0 \leq \eta \leq 2.5$), and neglect any bin to bin correlations, as well as any systematic uncertainties. 
We can thus define the $\lambda$ test statistic following Sec.~40 of ref.~\cite{Zyla:2020zbs} (see ref.~\cite{Cowan:1998ji, Cowan:2010js} for further details):
\begin{equation}
    -2 \log  \lambda(m_{\mathrm{LQ}},\beta) = -2 \log \left( \frac{\mathcal{P}(\mathbf{n};m_{\mathrm{LQ}},\beta)}{\mathcal{P}(\mathbf{n};\mathrm{SM})}\right) \, ,
    \label{eq:lambda_def}
\end{equation}
where $\mathbf{n}=(n_1, n_2, \cdots, n_{N_{\mathrm{bin}}})$ is the number of events observed in different $\eta$ bins (which we assumed is equal to the SM prediction), $\mathcal{P}(\mathbf{n};m_{\mathrm{LQ}},\beta)$ is the probability of seeing this experimental outcome assuming the underlying model is our LQ model with a given mass and Yukawa coupling, and $\mathcal{P}(\mathbf{n};\mathrm{SM})$ is the probability of finding the same experimental outcome from the SM. 
According to Wilks' theorem~\cite{Wilks:1938dza}, for large enough number of events, the test statistic $\lambda$ will approach a $\chi^2$ distribution with $N_{\mathrm{bin}}$ degrees of freedom. 
Assuming a Poisson distribution in each bin, we find
\begin{equation}
    -2 \log  \lambda(m_{\mathrm{LQ}},\beta) =  2 \sum_{i=1}^{i=N_{\mathrm{bin}}} \left[   \mu_i (m_{\mathrm{LQ}},\beta) - b_i + b_i \log \left( \frac{b_i}{\mu_i (m_{\mathrm{LQ}},\beta)}  \right)            \right]\, ,
    \label{eq:lambda_exclusion}
\end{equation}
where $b_i$ is the SM prediction for the number of events in each bin (and according to our assumption $b_i=n_i$ for this test), and $\mu_i (m_{\mathrm{LQ}},\beta)$ is the LQ model prediction for bin $i$'s event count. We calculate this test statistic for each point in the parameter space and find the expected $95\%$ confidence level exclusion bound from a \muc with various COM energies. 

In addition to the projected 95\% C.L. exclusion bounds, we also wish to calculate the $5\sigma$ discovery reach of the \muc.
To do so, we modify the calculation above by assuming the experimental outcome is distributed among the bins according to the prediction of our LQ model with a given mass and Yukawa coupling. 
With this assumption, we find for what parameters the observation would exclude the SM with greater than $5\sigma$ confidence. 
The test statistic we use for this case is thus
\begin{eqnarray}
    -2 \log  \lambda(m_{\mathrm{LQ}},\beta) &=& -2 \log \left( \frac{\mathcal{P}(\mathbf{n};\mathrm{SM})}{\mathcal{P}(\mathbf{n};m_{\mathrm{LQ}},\beta)}\right) \nonumber \\ 
    &=& 2 \sum_{i=1}^{i=N_{\mathrm{bin}}} \left[    b_i- \mu_i (m_{\mathrm{LQ}},\beta) + \mu_i (m_{\mathrm{LQ}},\beta) \log \left( \frac{\mu_i (m_{\mathrm{LQ}},\beta)}{b_i}  \right)            \right] \, ,
    \label{eq:lambda_def2}
\end{eqnarray}
where again we have assumed a Poisson distribution for number of events in each bin.

For the PP and SP channels, we take a simpler ``cut-and-count'' based approach, and first compute the expected number of background events after all selection cuts are taken into account.
For the $95\%$ C.L. constraints, we then assume the observed number of events is equal to the expected background and compute the upper limit of the 95\% C.L. Bayesian posterior probability, using the ``frequentist'' prior, i.e., we solve for $s_{\textrm{up}}$ in
\begin{equation}
0.95 = \frac{\int_{0}^{s_{\textrm{up}}} P(b | s) \dd s}{\int_0^{\infty} P(b | s) \dd s}
\end{equation}
where $b$ is the expected background, $s$ is the expected signal as a function of the model parameters, and $P(n | s)$ is the Poisson probability for observing $n$ events assuming a mean of $s + b$:
\begin{equation}
    P(n | s) = \frac{(s+b)^n}{n!} e^{-(s+b)} .
\end{equation}
See ref.~\cite{Cowan:1998ji, Cowan:2010js} for more details.
To compute the $5\sigma$ discovery reach, we follow a similar procedure, except that we instead solve for the number of observed events necessary to exclude the expected background, $b$ at $5\sigma$ confidence ($p = 5.7\times 10^{-7}$).
We checked that this approach leads to the same limits as obtained from the likelihood analysis described above with $N_{\textrm{bin}} = 1$ for the PP and SP channels.

\section{Modified Gauge Interactions}
\label{app:kappa}

As mentioned at the beginning of Sec.~\ref{sec:production}, non-zero values of $\kappa_U$, $\tilde{\kappa}_U$ can arise in different UV completions, but are otherwise ignored in this paper. In this appendix, we briefly outline how the phenomenology changes when the assumption that $\kappa_U$, $\tilde{\kappa}_U = 0$ is relaxed.

First, we note that the diagrams that interfere with the SM DY production process discussed in Sec.~\ref{subsec:DY} are completely insensitive to the strong and EW interactions of the $U_1$, and depend only on the couplings $g_U$ and the flavor spurions, $\beta_L^{i2}$.
Therefore, all of our DY results are unchanged in the non-standard gauge coupling scenario.

Pair-production and single-production of the LQ on the other hand, depend heavily on the electroweak interactions of the LQ. 
All three SP diagrams in Fig.~\ref{fig:diag_singleprod} depend on the gauge couplings of the $U_1$ and its bounds will get weaker for nonzero $\tilde{\kappa}_U$. Since the SP channels are already not very constraining in most of the LQ parameter space, we will not consider these modifications any further.

Thus, we are left to understand the importance of $\tilde{\kappa}_U$ on the PP constraints.
We recompute the bounds from pair production as in Sec.~\ref{subsec:PP}, with $\tilde{\kappa}_{U} = 0.5$ and $1.0$. The latter corresponds to the ``minimal coupling'' scenario, as defined in refs.~\cite{CiezaMontalvo:1992bs, Blumlein:1992ej}, though we emphasize that there are still non-vanishing electroweak interactions as required by gauge-invariance.

\begin{figure}
\centering
\includegraphics[width=0.45\linewidth]{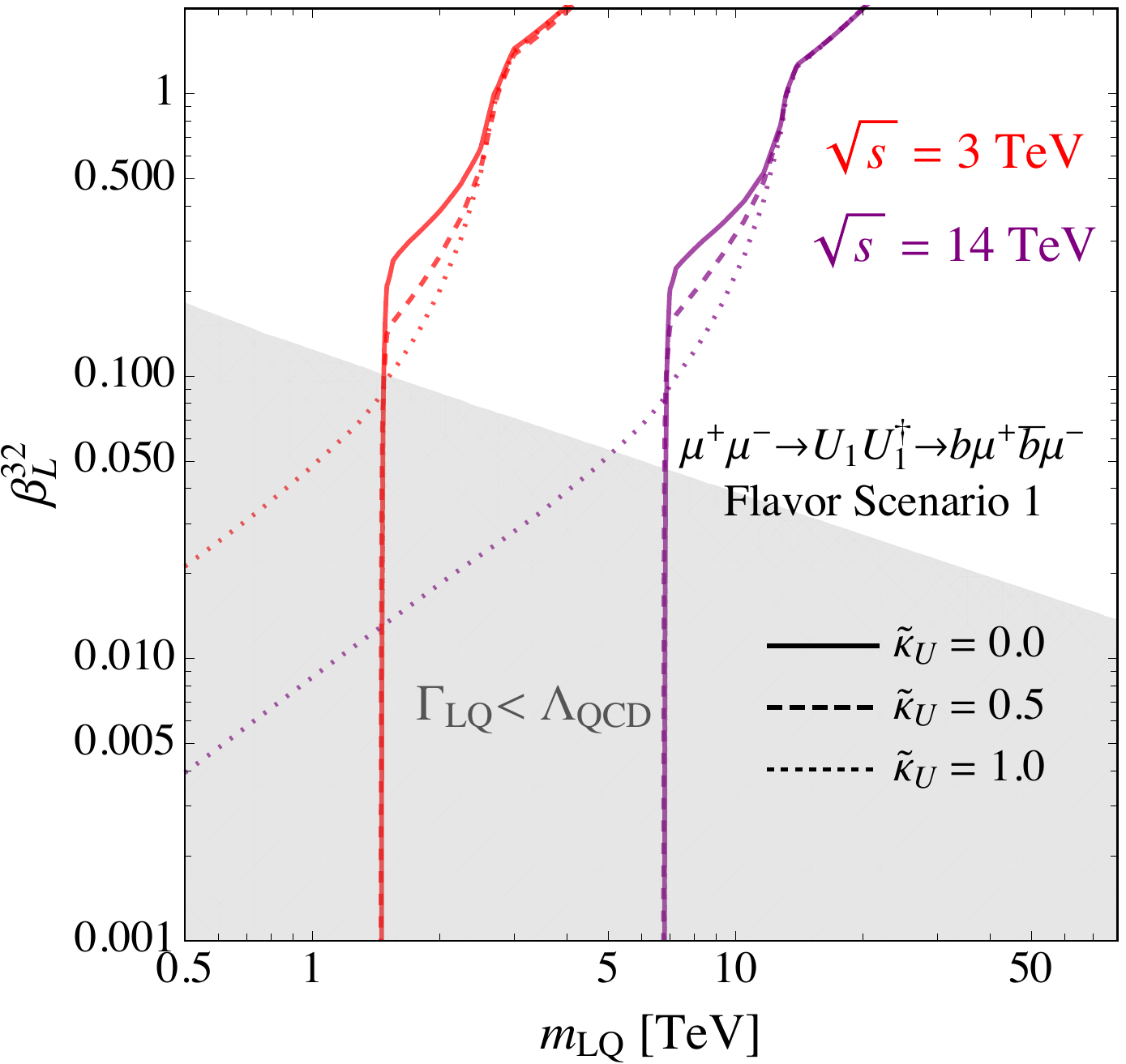}
~
\includegraphics[width=0.45\linewidth]{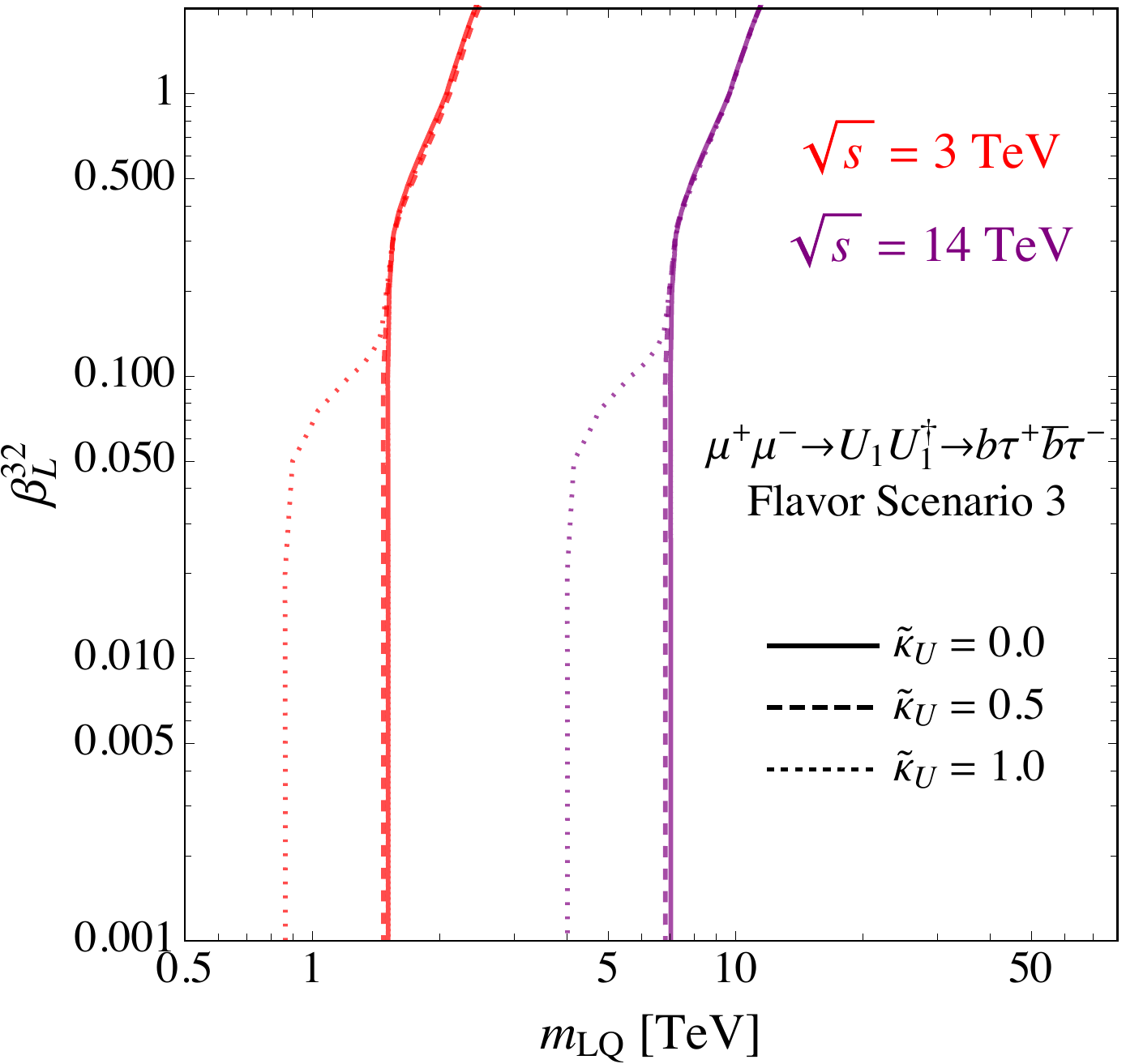}
\caption{
Contours showing the $5\sigma$ discovery reach of a $\sqrt{s} = 3$ or $14\,\textrm{TeV}$ muon collider via pair production for several values of the modified gauge coupling, governed by $\tilde{\kappa}_U$ (see Eq.~\eqref{eq:u1_lag}). The solid, dashed, and dotted lines indicate the reach with $\tilde{\kappa}_U = 0.0$, $0.5$, and $1.0$, respectively.
}\label{fig:vlqpair_contours_kappa}
\end{figure}

The resulting constraints for flavor scenarios 1 and 3 are shown in the left and right panels of Fig.~\ref{fig:vlqpair_contours_kappa}.
Here we show only the $5\sigma$ discovery reaches, for three different values of $\tilde{\kappa}_U$ at $\sqrt{s} = 3$ and $14\,\textrm{TeV}$.
We see that for very large values of $\beta_L^{32}$, the bounds are unchanged, as they are set by the $t$-channel production of the LQ pairs and insensitive to the gauge couplings. For smaller values of $\beta_L^{32}$, particularly in flavor scenario 1, we see that the bounds become substantially weaker, and for $\tilde{\kappa}_U = 1.0$, cannot reach all the way to the $\sqrt{s}/2$ threshold. At intermediate values, the destructive interference between the $t$- and $s$-channel diagrams in Fig.~\ref{fig:diag_pairprod} leads to larger values of $\tilde{\kappa}_U$ strengthening the bounds in flavor scenario 1. 
In flavor scenario 3, the smaller background makes the constraints less sensitive to the modified interactions, though the reach is somewhat decreased in the minimal coupling scenario to $\sim \sqrt{s}/3$, rather than $\sqrt{s}/2$. As a final note, we remind that these constraints do not use the full power of a resonance scan, so they could potentially still be improved, particularly in flavor scenario 1 where the background is non-negligible.

\bibliographystyle{utphys}
\bibliography{ref}

\end{document}